\documentclass[aps,prd,twocolumn,superscriptaddress]{revtex4}
\usepackage{epsfig,epsf}
\usepackage{amsmath}
\usepackage{amsthm}
\usepackage{amsfonts}
\usepackage{amssymb}
\usepackage{dsfont}
\usepackage{multirow}
\usepackage{appendix}
\usepackage{slashed}
\usepackage[active]{srcltx}
\usepackage{psfrag}

\begin{document}

\title{Hadronic molecule model for the doubly charmed state $T^{+}_{cc}$}
\date{\today}
\author{S.~S.~Agaev}
\affiliation{Institute for Physical Problems, Baku State University, Az--1148 Baku,
Azerbaijan}
\author{K.~Azizi}
\affiliation{Department of Physics, University of Tehran, North Karegar Avenue, Tehran
14395-547, Iran}
\affiliation{Department of Physics, Do\v{g}u\c{s} University, Dudullu-\"{U}mraniye, 34775
Istanbul, Turkey}
\author{H.~Sundu}
\affiliation{Department of Physics, Kocaeli University, 41380 Izmit, Turkey}

\begin{abstract}
The mass, current coupling, and width of the doubly charmed four-quark meson
$T_{cc}^{+}$ are explored by treating it as a hadronic molecule $%
M_{cc}^{+}\equiv D^{0}D^{\ast +}$. The mass and current coupling of this
molecule are calculated using the QCD two-point sum rule method by including
into analysis contributions of various vacuum condensates up to dimension $%
10 $. The prediction for the mass $m=(4060\pm 130)~\mathrm{MeV}$ exceeds the
two-meson $D^{0}D^{\ast +}$ threshold $3875.1~\mathrm{MeV}$, which makes
decay of the molecule $M_{cc}^{+}$ to a pair of conventional mesons $%
D^{0}D^{\ast +}$ kinematically allowed process. The strong coupling $G $ of
particles at the vertex $M_{cc}^{+}D^{0}D^{\ast +}$ is found by applying the
QCD three-point sum rule approach, and used to evaluate the width of the
decay $M_{cc}^{+} \to D^{0}D^{\ast +}$. Obtained result for the width $%
\Gamma=(3.8\pm 1.7)~\mathrm{MeV}$ demonstrates that $M_{cc}^{+}$ is wider
than the resonance $T_{cc}^{+}$.
\end{abstract}

\maketitle


\section{Introduction}

\label{sec:Int}
Recently, the LHCb collaboration informed about observation, for the first
time, of a doubly charmed axial-vector state $T_{cc}^{+}$ \ composed of four
quarks $cc\overline{u}\overline{d}$ \cite{Aaij:2021vvq,LHCb:2021auc}. This
state was fixed in $D^{0}D^{0}\pi ^{+}$ mass distribution as a narrow peak
with the width $\Gamma =(410\pm 165\pm 43_{-38}^{+18})~\mathrm{keV}$, which
means that it is longest living exotic meson discovered till now. The mass
of $T_{cc}^{+}$ is very close to the two-meson $D^{0}D^{\ast +}$ threshold $%
3875.1~\mathrm{MeV}$, but is smaller than this limit by an amount of $\delta
m_{\exp }=(-273\pm 61\pm 5_{-14}^{+11})~\mathrm{keV}$. These features of $%
T_{cc}^{+}$, in particular its narrow width, made the doubly charmed exotic
meson $T_{cc}^{+}$ an object of intensive studies \cite%
{Agaev:2021vur,Feijoo:2021ppq,Yan:2021wdl,Fleming:2021wmk,
Azizi:2021aib,Meng:2021jnw,Ling:2021bir,Chen:2021vhg,Xin:2021wcr}.

It is worth emphasizing that doubly charmed tetraquarks attracted already
interests of researchers. This is connected with estimated stability some of
tetraquarks containing heavy diquarks $bb$, $bc$ and $cc$ against strong and
maybe electromagnetic decays. If exist, such particles can transform to
mesons only through weak decays, and have mean lifetimes which would be
considerably longer than that of conventional mesons \cite%
{Agaev:2018khe,Agaev:2019lwh,Agaev:2020zag,Agaev:2019kkz}. There is growing
conviction that tetraquarks built of $bb$ diquarks are stable particles,
whereas the situation with ones composed of $bc$ and $cc$ diquarks is still
remaining controversial \cite{Karliner:2017qjm,Eichten:2017ffp,Agaev:2020zad}%
.

Because the present work is devoted to investigation of doubly charmed
states, below we restrict ourselves by analyses of problems and achievements
connected only with these particles. Thus, tetraquarks $cc\overline{q}%
\overline{q}^{\prime }$ were theoretically studied using different methods
of the high energy physics. In the framework of the QCD sum rule method they
were analyzed in Refs.\ \cite{Navarra:2007yw,Du:2012wp}. In the first
article the authors explored the axial-vector tetraquark $cc\overline{u}%
\overline{d}$. Prediction for its mass $(4000\pm 200)~\mathrm{MeV}$ implies
that the axial-vector tetraquark $cc\overline{u}\overline{d}$ is unstable
and readily decays to mesons $D^{0}D^{\ast }{}^{+}$. Four-quark exotic
mesons of general $cc\overline{q}\overline{q}^{\prime }$ content and quantum
numbers $J^{\mathrm{P}}=0^{-},~0^{+},~1^{-}$ and $1^{+}$ were investigated
in Ref.\ \cite{Du:2012wp}. In accordance with results of this analysis,
masses of tetraquarks $cc\overline{q}\overline{q}$, $cc\overline{q}\overline{%
s}$, and $cc\overline{s}\overline{s}$ are above corresponding thresholds for
all explored quantum numbers. In other words, a class of tetraquarks
composed of a diquark $cc$ and a light antidiquark does not contain
strong-interaction stable particles.

Discovery of doubly charmed baryon $\Xi _{cc}^{++}=ccu$ by the LHCb
collaboration \cite{Aaij:2017ueg}, and extracted experimental information
stimulated relatively new studies of heavy tetraquarks. A reason was that,
these experimental data were employed as new input parameters in a
phenomenological model to estimate masses of the axial-vector tetraquarks $%
T_{bb;\overline{u}\overline{d}}^{-}$ and $T_{cc;\overline{u}\overline{d}%
}^{+} $ \cite{Karliner:2017qjm,Eichten:2017ffp}. In these articles it was
demonstrated that $T_{cc;\overline{u}\overline{d}}^{+}$ has the mass $%
(3882\pm 12)~\mathrm{MeV}$ and $3978~\mathrm{MeV}$, respectively, which are
above thresholds for both $D^{0}D^{\ast +}$ and $D^{0}D^{+}\gamma $ decays.
Other members of $cc\overline{q}\overline{q}$, and $cc\overline{q}\overline{s%
}$ families were considered in Ref.\ \cite{Eichten:2017ffp}: None of them
were classified as a stable state. Similar conclusions about properties of $%
T_{cc;\overline{u}\overline{d}}^{+}$ were drawn in Refs.\ \cite%
{Wang:2017dtg,Braaten:2020nwp,Cheng:2020wxa} as well. Contrary to these
studies, in Ref.\ \cite{Meng:2020knc} the authors calculated the mass of $%
T_{cc;\overline{u}\overline{d}}^{+}$ using a constituent quark model and
found that it is $23~\mathrm{MeV}$ below the two-meson threshold. Stable
nature of $T_{cc;\overline{u}\overline{d}}^{+}$ was demonstrated also by
means of lattice simulations \cite{Junnarkar:2018twb}, in which its mass was
estimated $(-23\pm 11)~\mathrm{MeV}$ below the two-meson threshold.

Detailed studies of pseudoscalar and scalar exotic mesons $cc\overline{u}%
\overline{d}$ were done in Ref.\ \cite{Agaev:2019qqn}. Analysis performed
there, demonstrated that these particles are strong-interaction unstable
structures, and fall apart to conventional mesons. Full widths of these
tetraquarks were evaluated by utilizing their decays to $D^{+}D^{\ast
}(2007)^{0}$, $D^{0}D^{\ast }(2010)^{+}$, and $D^{0}D^{+}$ mesons,
respectively. It was found, that these structures with widths $\sim 130~%
\mathrm{MeV}$ and $\sim 12~\mathrm{MeV}$ are relatively wide resonances.
Structures $cc\overline{s}\overline{s}$ and $cc\overline{d}\overline{s}$
form another interesting subgroup of doubly charmed tetraquarks, because
they are also doubly charged particles. Masses and widths of such
pseudoscalar tetraquarks were evaluated in Ref.\ \cite{Agaev:2018vag}.

Doubly charmed four-quark structures were studied also in the context of the
hadronic molecule picture, i.e., they were modeled as molecules of
conventional mesons. It is worth noting that charmonium molecules are not
new objects for investigations: Problems of such compounds were addressed in
literature decades ago \cite{Novikov:1977dq}. As a hadronic molecule $%
M_{cc}^{+}\equiv D^{0}D^{\ast +}$ built of ordinary mesons $D^{0}$ and $%
D^{\ast +}$, the axial-vector state $cc\overline{u}\overline{d}$ was
considered in Refs.\ \cite{Dias:2011mi,Li:2012ss}. The mass of $M_{cc}^{+}$
was estimated in Ref.\ \cite{Dias:2011mi} using the QCD spectral sum rule
approach. Obtained prediction $(3872.2\pm 39.5)~\mathrm{MeV}$ shows that
this molecule cannot decay to mesons $D^{0}$ and $D^{\ast +}$, but its mass
is enough to trigger the strong decay $M_{cc}^{+}\rightarrow D^{0}D^{0}\pi
^{+}$.

In our recent article, we treated $T_{cc}^{+}$ as an axial-vector
diquark-antidiquark (tetraquark) state with quark content $cc\overline{u}%
\overline{d}$, and calculated its spectroscopic parameters and full width
\cite{Agaev:2021vur}. Computations performed in the context of the QCD
two-point sum rule method led for the mass of this state to the result $%
(3868\pm 124)~\mathrm{MeV}$, which is consistent with the LHCb measurements.
This means that $T_{cc}^{+}$ does not decay to a meson pair $D^{0}D^{\ast +}$%
. Therefore, we evaluated full width of $T_{cc}^{+}$ by considering its
alternative strong decay channels. In fact, production of $D^{0}D^{0}\pi
^{+} $ can run through decay of $T_{cc}^{+}$ to a scalar tetraquark $T_{cc;%
\overline{u}\overline{u}}^{0}$ and $\pi ^{+}$ followed by the process $T_{cc;%
\overline{u}\overline{u}}^{0}\rightarrow D^{0}D^{0}$. The process $%
T_{cc}^{+}\rightarrow \widetilde{T}\pi ^{0}\rightarrow D^{0}D^{+}\pi ^{0}$
is another decay mode of $T_{cc}^{+}$. Here, $\widetilde{T}$ \ is the scalar
exotic meson with content $cc\overline{u}\overline{d}$. This means, that in
our analysis decays to scalar tetraquarks $T_{cc;\overline{u}\overline{u}%
}^{0}$ and $\widetilde{T}$ \ was considered as a dominant mechanism for
transformation of $T_{cc}^{+}$ . Full width of $T_{cc}^{+}$ estimated in
Ref.\ \cite{Agaev:2021vur} equals to $\Gamma =(489\pm 92)~\mathrm{keV}$
which nicely agrees with the experimental data.

As is seen, an assumption about the diquark-antidiquark structure of $%
T_{cc}^{+}$ gives for its mass and width results compatible with the LHCb
data \cite{Agaev:2021vur}. In accordance to Ref.\ \cite{Dias:2011mi}, the
molecule model for the mass of $M_{cc}^{+}$ leads to almost the same
prediction. Unfortunately, in this paper the authors did not compute width
of the molecule $M_{cc}^{+}$, therefore it is difficult to declare a full
convergence of results for $T_{cc}^{+}$ and $M_{cc}^{+}$ obtained in the
framework of the QCD sum rule method. The reason is that masses of $%
T_{cc}^{+}$ and $M_{cc}^{+}$ were extracted, as usual, with theoretical
uncertainties, and due to overlapping of relevant regions, this information
is not enough to distinguish diquark-antidiquark and molecule states. To
make reliable statements about internal organization of the four-quark state
seen by LHCb, it is necessary to investigate decay modes of this particle,
and calculate its full width.

The program outlined above was realized in the diquark-antidiquark picture
in our article \cite{Agaev:2021vur}. In the present work, we consider this
problem in the framework of the hadronic molecule model, and calculate the
mass and width of $M_{cc}^{+}$. We wish to answer a question whether both
the mass and width of $M_{cc}^{+}$ agree with new LHCb data. For these
purposes, we calculate the spectroscopic parameters of $M_{cc}^{+}$ using
the QCD two-point sum rule method \cite{Shifman:1978bx,Shifman:1978by}. Our
analysis proves that the mass of $M_{cc}^{+}$ exceeds the LHCb data, which
makes the process $M_{cc}^{+}\rightarrow D^{0}D^{\ast +}$ kinematically
allowed one. The width of this decay channel is found by means of the
three-point version of QCD sum rule approach: It is used to extract the
strong coupling $G$ at the vertex $M_{cc}^{+}D^{0}D^{\ast +}$.

This article is structured in the following manner: In Sec.\ \ref{sec:Mass},
we compute the mass $m$ and coupling $f$ of the molecule $M_{cc}^{+}$ in the
context of the QCD two-point sum rule method. In these calculations, we take
into account various vacuum condensates up to dimension $10$. In Sec.\ \ref%
{sec:Width}, we consider the decay mode $M_{cc}^{+}\rightarrow D^{0}D^{\ast
+}$, find the strong coupling $G$ and evaluate the width of this process. We
reserve Sec.\ \ref{sec:Conclusion} for discussion and conclusions.


\section{Spectroscopic parameters of $M_{cc}^{+}$}

\label{sec:Mass}

The sum rules necessary to evaluate the spectroscopic parameters of the
molecule $M_{cc}^{+}$ can be derived from analysis of the correlation
function%
\begin{equation}
\Pi _{\mu \nu }(p)=i\int d^{4}xe^{ipx}\langle 0|\mathcal{T}\{J_{\mu
}(x)J_{\nu }^{\dag }(0)\}|0\rangle,  \label{eq:CF1}
\end{equation}%
where $J_{\mu }(x)$ is the interpolation current for the axial-vector state $%
M_{cc}^{+}$. In the hadronic molecule model the current $J_{\mu }(x)$ is
given by the expression
\begin{equation}
J_{\mu }(x)=\overline{d}_{a}(x)\gamma _{\mu }c_{a}(x)\overline{u}%
_{b}(x)\gamma _{5}c_{b}(x),  \label{eq:Curr1}
\end{equation}%
where $a$ and $b$ are color indices.

To find the sum rules for $m$ and $f$, we express the correlation function $%
\Pi _{\mu \nu }(p)$ in terms of $M_{cc}^{+}$ molecule's physical parameters.
Because $M_{cc}^{+}$ is composed of ground-state mesons $D^{0}$ and $D^{\ast
+}$, it can be treated as lowest lying system in this class of particles.
Therefore, in the correlation function $\Pi _{\mu \nu }^{\mathrm{Phys}}(p)$,
we write down explicitly only first term that corresponds to $M_{cc}^{+}$
\begin{equation}
\Pi _{\mu \nu }^{\mathrm{Phys}}(p)=\frac{\langle 0|J_{\mu
}|M_{cc}^{+}(p,\epsilon )\rangle \langle M_{cc}^{+}(p,\epsilon )|J_{\nu
}^{\dagger }|0\rangle }{m^{2}-p^{2}}+\cdots .  \label{eq:PhysSide}
\end{equation}%
The $\Pi _{\mu \nu }^{\mathrm{Phys}}(p)$ is obtained by inserting into the
correlation function Eq.\ (\ref{eq:CF1}) the full set of states with
spin-parities $J^{\mathrm{P}}=1^{+}$, and carrying out integration over $x$.
The dots in Eq.\ (\ref{eq:PhysSide}) denote contributions coming from higher
resonances and continuum states.

To derive Eq.\ (\ref{eq:PhysSide}), we assume that the physical side of the
sum rule can be approximated by a single pole term. In the case of the
multiquark systems $\Pi _{\mu \nu }^{\mathrm{Phys}}(p)$ receives
contribution, however, also from two-meson reducible terms \cite%
{Kondo:2004cr,Lee:2004xk}. That is because the current $J_{\mu }(x)$
interacts not only with a molecule $M_{cc}^{+}$, but also with the two-meson
continuum with the same quantum numbers and quark content. Effects of
current-continuum interaction, properly taken into account, generates a
finite width $\Gamma (p^{2})$ of the hadronic molecule and leads to the
modification in Eq.\ (\ref{eq:PhysSide}) in accordance with the prescription
\cite{Wang:2015nwa}
\begin{equation}
\frac{1}{m^{2}-p^{2}}\rightarrow \frac{1}{m^{2}-p^{2}-i\sqrt{p^{2}}\Gamma
(p^{2})}.  \label{eq:Modification}
\end{equation}%
The two-meson contributions can be included into analysis by rescaling the
coupling $f$ of $M_{cc}^{+}$, and keeping untouched its mass. Calculations
demonstrated that these effects are small and do not exceed uncertainties of
sum rule calculations. Indeed, in the case of the doubly charmed
pseudoscalar tetraquark $cc\overline{s}\overline{s}$ with the mass $%
m_{T}=4390~\mathrm{MeV}$ and full width $\Gamma _{T}\approx 300~\mathrm{MeV}$%
, two-meson effects lead to additional $\approx 7\%$ uncertainty in the
current coupling $f_{T}$ \cite{Agaev:2018vag}. For the resonance $%
Z_{c}^{-}(4100)$ these ambiguities amount to $\approx 5\%$ of the coupling $%
f_{Z_{c}}$ \cite{Sundu:2018nxt}. As we shall see below, the molecule $%
M_{cc}^{+}$ has the width $(3.8\pm 1.7)~\mathrm{MeV}$. Therefore,
aforementioned effects are negligible, and in $\Pi _{\mu \nu }^{\mathrm{Phys}%
}(p)$ it is enough to employ the zero-width single-pole approximation.

The function $\Pi _{\mu \nu }^{\mathrm{Phys}}(p)$ can be presented in a more
compact form. To this end, we introduce the matrix element
\begin{equation}
\langle 0|J_{\mu }|M_{cc}^{+}(p,\epsilon )\rangle =fm\epsilon _{\mu },
\label{eq:MElem1}
\end{equation}%
where $\epsilon _{\mu }$ is the polarization vector of the molecule $%
M_{cc}^{+}$. It is not difficult to demonstrate that the function $\Pi _{\mu
\nu }^{\mathrm{Phys}}(p)$ in terms of $m$ and $f$ \ has the simple form
\begin{equation}
\Pi _{\mu \nu }^{\mathrm{Phys}}(p)=\frac{m^{2}f^{2}}{m^{2}-p^{2}}\left(
-g_{\mu \nu }+\frac{p_{\mu }p_{\nu }}{m^{2}}\right) +\cdots .
\label{eq:PhysSide1}
\end{equation}

The QCD side of the sum rules $\Pi _{\mu \nu }^{\mathrm{OPE}}(p)$ has to be
calculated in the operator product expansion ($\mathrm{OPE}$) with some
fixed accuracy. To find $\Pi _{\mu \nu }^{\mathrm{OPE}}(p)$, we calculate
the correlation function using explicit form of the current $J_{\mu }(x)$.
As a result, we express $\Pi _{\mu \nu }^{\mathrm{OPE}}(p)$ in terms of
heavy and light quark propagators
\begin{eqnarray}
&&\Pi _{\mu \nu }^{\mathrm{OPE}}(p)=i\int d^{4}xe^{ip\cdot x}\left\{ \mathrm{%
Tr}\left[ \gamma _{5}S_{c}^{bb^{\prime }}(x)\gamma _{5}S_{u}^{b^{\prime
}b}(-x)\right] \right.  \notag \\
&&\times \mathrm{Tr}\left[ \gamma _{\mu }S_{c}^{aa^{\prime }}(x)\gamma _{\nu
}S_{d}^{a^{\prime }a}(-x)\right] -\mathrm{Tr}\left[ \gamma _{\mu
}S_{c}^{ab^{\prime }}(x)\gamma _{5}\right.  \notag \\
&&\left. \left. \times S_{u}^{b^{\prime }b}(-x)\gamma _{5}S_{c}^{ba^{\prime
}}(x)\gamma _{\nu }S_{d}^{a^{\prime }a}(-x)\right] \right\} .
\label{eq:QCDSide}
\end{eqnarray}%
In Eq.\ (\ref{eq:QCDSide}) $S_{q}^{ab}(x)$ and $S_{c}^{ab}(x)$ are
propagators of $q(u,d)$ and $c$-quarks, formulas for which are collected in
Appendix.

The QCD sum rules can be derived using the same Lorentz structures in $\Pi
_{\mu \nu }^{\mathrm{Phys}}(p)$ and $\Pi _{\mu \nu }^{\mathrm{OPE}}(p)$. For
our purposes, the structures proportional to $g_{\mu \nu }$ are appropriate,
because they are free of contributions of spin-$0$ particles. To obtain a
sum rule, we equate invariant amplitudes $\Pi ^{\mathrm{Phys}}(p^{2})$ and $%
\Pi ^{\mathrm{OPE}}(p^{2})$ corresponding to these structures, and apply the
Borel transformation to both sides of the obtained expression. The last
operation is necessary to suppress contributions stemming from the higher
resonances and continuum states. At the following phase of manipulations, we
make use an assumption about the quark-hadron duality, and subtract\ from
the physical side of the equality higher resonances' and continuum
contributions. By this way, the final sum rule equality acquires a
dependence on the Borel $M^{2}$ and continuum threshold (subtraction) $s_{0}$
parameters. This equality, and second expression obtained by applying the
operator $d/d(-1/M^{2})$ to its both sides, form a system which is used to
find sum rules for the mass $m$ and coupling $f$
\begin{equation}
m^{2}=\frac{\Pi ^{\prime }(M^{2},s_{0})}{\Pi (M^{2},s_{0})},  \label{eq:Mass}
\end{equation}%
\begin{equation}
f^{2}=\frac{e^{m^{2}/M^{2}}}{m^{2}}\Pi (M^{2},s_{0}),  \label{eq:Coupling}
\end{equation}%
where $\Pi ^{\prime }(M^{2},s_{0})=d\Pi (M^{2},s_{0})/d(-1/M^{2})$.

In Eqs.\ (\ref{eq:Mass}) and (\ref{eq:Coupling}) the function $\Pi
(M^{2},s_{0})$ is Borel transformed and continuum subtracted invariant
amplitude $\Pi ^{\mathrm{OPE}}(p^{2})$. We calculate $\Pi (M^{2},s_{0})$ by
taking into account quark, gluon and mixed vacuum condensates up to
dimension $10$. It has the following form
\begin{equation}
\Pi (M^{2},s_{0})=\int_{4m_{c}^{2}}^{s_{0}}ds\rho ^{\mathrm{OPE}%
}(s)e^{-s/M^{2}}+\Pi (M^{2}),  \label{eq:InvAmp}
\end{equation}%
where $\rho ^{\mathrm{OPE}}(s)$ is the two-point spectral density. The
second component of the invariant amplitude $\Pi (M^{2})$ contains
nonperturbative contributions calculated directly from $\Pi _{\mu \nu }^{%
\mathrm{OPE}}(p)$. The explicit expression of the function $\Pi
(M^{2},s_{0}) $ is removed to Appendix.

The quark, gluon and mixed condensates which enter to the sum rules (\ref%
{eq:Mass}) and (\ref{eq:Coupling}) are universal parameters of computations:
\begin{eqnarray}
&&\langle \overline{q}q\rangle =-(0.24\pm 0.01)^{3}~\mathrm{GeV}^{3},\
\notag \\
&&\langle \overline{q}g_{s}\sigma Gq\rangle =m_{0}^{2}\langle \overline{q}%
q\rangle ,\ m_{0}^{2}=(0.8\pm 0.1)~\mathrm{GeV}^{2},\   \notag \\
&&\langle \frac{\alpha _{s}G^{2}}{\pi }\rangle =(0.012\pm 0.004)~\mathrm{GeV}%
^{4},  \notag \\
&&\langle g_{s}^{3}G^{3}\rangle =(0.57\pm 0.29)~\mathrm{GeV}^{6},  \notag \\
&&m_{c}=1.275\pm 0.025~\mathrm{GeV}.  \label{eq:Parameters}
\end{eqnarray}%
The correlation function $\Pi (M^{2},s_{0})$ depends on the $c$ quark mass,
numerical value of which is shown in Eq.\ (\ref{eq:Parameters}) as well.
Contrary, the Borel and continuum threshold parameters $M^{2}$ and $s_{0}$
are auxiliary quantities of calculations: Their choice depends on the
problem under consideration, and has to meet restrictions imposed on the
pole contribution ($\mathrm{PC}$) and convergence of $\mathrm{OPE}$.

To estimate the $\mathrm{PC}$, we use the expression
\begin{equation}
\mathrm{PC}=\frac{\Pi (M^{2},s_{0})}{\Pi (M^{2},\infty )}.  \label{eq:Pole}
\end{equation}%
The convergence of the operator product expansion is checked by means of the
formula
\begin{equation}
R(M^{2})=\frac{\Pi ^{\mathrm{DimN}}(M^{2},s_{0})}{\Pi (M^{2},s_{0})},
\label{eq:Convergence}
\end{equation}%
where $\Pi ^{\mathrm{DimN}}(M^{2},s_{0})$ is the contribution of the last
three terms in $\mathrm{OPE}$, i.e., $\mathrm{DimN=Dim(8+9+10)}$ .

In the current investigation, we use a restriction $\mathrm{PC\geq 0.2}$
which is typical for multiquark hadrons. We also consider $\mathrm{OPE}$ as
a convergent provided at the minimum of the Borel parameter the ratio $%
R(M^{2})$ is less than $0.01$. Calculations confirm that the working windows
that satisfy these requirements are
\begin{equation}
M^{2}\in \lbrack 4,6]~\mathrm{GeV}^{2},\ s_{0}\in \lbrack 19.5,21.5]~\mathrm{%
GeV}^{2}.  \label{eq:Regions}
\end{equation}%
In fact, within these regions $\mathrm{PC}$ changes on average in limits $%
0.20\leq \mathrm{PC}\leq 0.61,$ and at the minimum $M^{2}=4~\mathrm{GeV}^{2}$%
, we get $R(M^{2})\leq 0.01$. In general, sum rules' predictions should not
depend on the choice of $M^{2}$, but in real analysis there is a undesirable
dependence of $m$ and $f$ on the Borel parameter $M^{2}$. Therefore, the
window for $M^{2}$ should minimize this dependence as well, and the region
from Eq.\ (\ref{eq:Regions}) obeys this condition.

To extract the mass $m$ and coupling $f$, we calculate them at different
choices of the parameters $M^{2}$ and $s_{0}$, and find their values
averaged over the working regions Eq.\ (\ref{eq:Regions})
\begin{eqnarray}
m &=&(4060\pm 130)~\mathrm{MeV},  \notag \\
f &=&(5.1\pm 0.8)\times 10^{-3}~\mathrm{GeV}^{4}.  \label{eq:Result1}
\end{eqnarray}%
\ These results correspond to the point $M^{2}=5~\mathrm{GeV}^{2}$ and $%
s_{0}=20.4~\mathrm{GeV}^{2}$ which is approximately at middle of the regions
Eq.\ (\ref{eq:Regions}). The pole contribution computed at this point is
equal to $\mathrm{PC}\approx 0.53$, which guarantees credibility of obtained
predictions, and the ground-state nature of $M_{cc}^{+}$ in its class of
particles.

\begin{figure}[h]
\includegraphics[width=8.5cm]{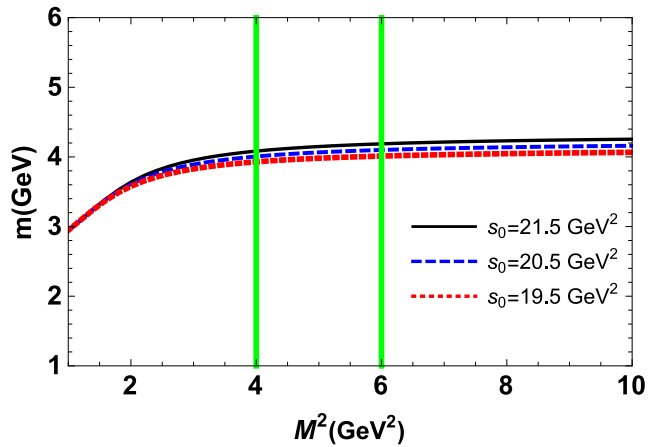}
\caption{The mass of the hadronic molecule $M_{cc}^{+}$ as a function of the
Borel parameter $M^{2}$ at fixed $s_{0}$. Vertical lines show boundaries of
working region for $M^{2}$ used in numerical computations. }
\label{fig:Mass}
\end{figure}

The mass $m$ of the molecule $M_{cc}^{+}$ as a function of $M^{2}$ is
plotted in Fig.\ \ref{fig:Mass}. Here, we show dependence of $m$ on the
Borel parameter in a wide range of $M^{2}$. One can see, that predictions
obtained at values of $M^{2}$ from Eq.\ (\ref{eq:Regions}) are relatively
stable, though residual effects of $M^{2}$ on $m$ is evident in this region
as well: This is unavoidable feature of the sum rule method which limits its
accuracy. At the same time, this method allows one to estimate ambiguities
of performed analysis which is the case only for some of nonperturbative QCD
approaches.

\begin{figure}[h]
\includegraphics[width=8.5cm]{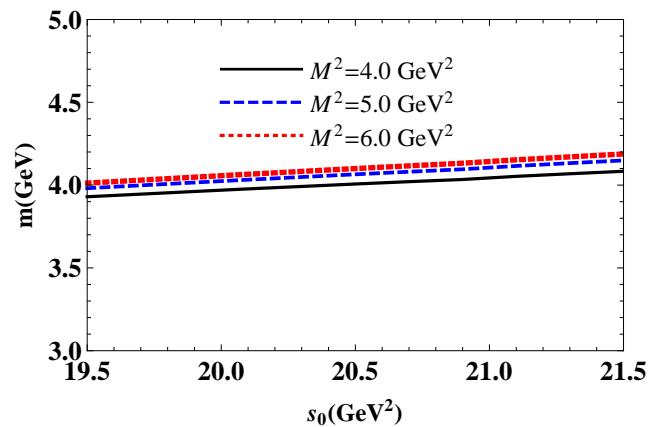}
\caption{Dependence of $m$ on the continuum threshold parameter $s_0$ at
fixed $M^2$. }
\label{fig:MassS}
\end{figure}

The second source of uncertainties is the continuum threshold parameter $%
s_{0}$, that separates a ground-state term from contributions of higher
resonances and continuum states. It carries also physical information about
first excitation of the $M_{cc}^{+}$, meaning that $\sqrt{s_{0}}$ should be
smaller than a mass of such state. Parameters of excited conventional
hadrons are known from theoretical studies or were measured in numerous
experiments. Therefore, a choice of the scale $s_{0}$ in relevant studies
does not create new problems. The mass spectra of multiquark hadrons, in
general, may have more complex structure. Additionally, there are only a few
resonances, which can be considered as radially or orbitally excited exotic
hadrons. Thus, the resonances $Z_{c}(3900)$ and $Z_{c}(4430)$ with a mass
gap $\approx 530~\mathrm{MeV}$ may be treated as the ground-state and first
radially excited axial-vector tetraquark $[cu][\overline{c}\overline{d}]$,
respectively \cite{Maiani:2014}. This conjecture was later confirmed by the
sum rule calculations in Refs.\ \cite{Wang:2014vha,Agaev:2017tzv}. The mass
spectra of the doubly heavy tetraquarks were analyzed in Ref.\ \cite%
{Kim:2022mpa} in the framework of a chiral-diquark picture. The difference
between masses of doubly charmed $1S$ and $2S$ axial-vector tetraquarks was
found there equal to $\approx 400~\mathrm{MeV}$. In light of these
investigations, $\sqrt{s_{0}}$ may exceed $m$ approximately $0.4-0.6~\mathrm{%
MeV}$. In our case, this gap is $\sqrt{s_{0}}-m\approx 450~\mathrm{MeV}$
which is a reasonable estimate for an exotic state composed of mesons $D^{0}$
and $D^{\ast +}$ and containing two $c$ quarks. Dependence of $m$ on the
continuum threshold parameter $s_{0}$ is shown in Fig.\ \ref{fig:MassS}.

The central value of the mass $m$ is above the two-meson $D^{0}D^{\ast
}(2010)^{+}$ threshold $3875.1~\mathrm{MeV}$, and exceeds the datum of the
LHCb collaboration. Even the low estimate for the mass $3930~\mathrm{MeV}$
overshoots this boundary. In other words, the dominant decay channel of the
molecule $M_{cc}^{+}$ is the process $M_{cc}^{+}\rightarrow D^{0}D^{\ast +}$%
. In the next section, we are going to calculate the width of $M_{cc}^{+}$
using this decay.


\section{Width of the decay $M_{cc}^{+}\rightarrow D^{0}D^{\ast +}$}

\label{sec:Width}

The four-quark state $T_{cc}^{+}$ was observed in $D^{0}D^{0}\pi ^{+}$ mass
distribution, and therefore it decays strongly to these mesons. The hadronic
molecule $M_{cc}^{+}$ has the same quantum numbers and quark content,
therefore the process $M_{cc}^{+}\rightarrow D^{0}D^{0}\pi ^{+}$ is among
possible decay modes of $M_{cc}^{+}$. This decay may proceed through two
stages: the process $M_{cc}^{+}\rightarrow D^{0}D^{\ast +}$ followed by the
decay $D^{\ast +}\rightarrow D^{0}\pi ^{+}$. Our calculations show that the
mass of $M_{cc}^{+}$ is enough to generate this chain of transformations.

In this section, we study the decay $M_{cc}^{+}\rightarrow D^{0}D^{\ast +}$
and find the strong coupling $G$ of particles at the vertex $%
M_{cc}^{+}D^{0}D^{\ast +}$. The QCD three-point sum rule for this coupling
can be derived from analysis of the correlation function
\begin{eqnarray}
&&\Pi _{\mu \nu }(p,p^{\prime })=i^{2}\int d^{4}xd^{4}ye^{i(p^{\prime
}y-px)}\langle 0|\mathcal{T}\{J_{\nu }^{D^{\ast }}(y)  \notag \\
&&\times J^{D}(0)J_{\mu }^{\dagger }(x)\}|0\rangle ,  \label{eq:CF2}
\end{eqnarray}%
where $J_{\mu }(x)$, $\ J_{\nu }^{D^{\ast }}(x)$ and $J^{D}(x)$ are the
relevant interpolating currents. For the molecule $M_{cc}^{+}$ the current $%
J_{\mu }(x)$ is given by Eq.\ (\ref{eq:Curr1}). The $J_{\nu }^{D^{\ast }}(x)$
and $J^{D}(x)$ are currents of the mesons $D^{\ast +}$ and $D^{0}$, which
have the following forms
\begin{equation}
J_{\nu }^{D^{\ast }}(x)=\overline{d}_{i}(x)\gamma _{\nu }c_{i}(x),\ J^{D}(x)=%
\overline{u}_{j}(x)i\gamma _{5}c_{i}(x),  \label{eq:Curr5}
\end{equation}%
where $i$ and $j$ are color indices. The 4-momenta of the particles $%
M_{cc}^{+}$ and $D^{\ast +}$ are $p$ and $p^{\prime }$, respectively, hence
the momentum of the $D^{0}$ meson is $q=p-p^{\prime }$.

We continue using standard prescriptions of the sum rule method and, first
calculate the correlation function $\Pi _{\mu \nu }(p,p^{\prime })$ in terms
of physical parameters of involved particles. Isolating in Eq.\ (\ref{eq:CF2}%
) \ a contribution of the ground-state particles, we get
\begin{eqnarray}
&&\Pi _{\mu \nu }^{\mathrm{Phys}}(p,p^{\prime })=\frac{\langle 0|J_{\nu
}^{D^{\ast }}|D^{\ast +}(p^{\prime },\varepsilon )\rangle \langle
0|J^{D}|D^{0}(q)\rangle }{(p^{2}-m^{2})(p^{\prime 2}-m_{D^{\ast
}}^{2})(q^{2}-m_{D}^{2})}  \notag \\
&&\times \langle M_{cc}^{+}(p,\epsilon )|J_{\mu }^{\dagger }|0\rangle
\langle D^{0}(q)D^{\ast +}(p^{\prime },\varepsilon )|M_{cc}^{+}(p,\epsilon
)\rangle +\cdots ,  \notag \\
&&  \label{eq:CF3}
\end{eqnarray}%
which is the physical side of the sum rule. In Eq.\ (\ref{eq:CF3}) $%
m_{D^{\ast }}$ and $m_{D}$ are the masses of the $D^{\ast +}$ and $D^{0}$
mesons \cite{PDG:2020}, respectively:
\begin{eqnarray}
m_{D^{\ast }} &=&(2010.26\pm 0.05)~\mathrm{MeV},  \notag \\
m_{D} &=&(1864.84\pm 0.05)~\mathrm{MeV}.  \label{eq:Masses}
\end{eqnarray}

For our purposes, it is necessary to employ the $D^{\ast +}$ and $D^{0}$
mesons' \ matrix elements, and, by this way to find compact expression for
the function $\Pi _{\mu \nu }^{\mathrm{Phys}}(p,p^{\prime })$.\ This can be
achieved by using the matrix elements
\begin{eqnarray}
&&\langle 0|J_{\nu }^{D^{\ast }}|D^{\ast +}(p^{\prime },\varepsilon )\rangle
=f_{D^{\ast }}m_{D^{\ast }}\varepsilon _{\nu },\   \notag \\
&&\langle 0|J^{D}|D^{0}(q)\rangle =\frac{f_{D}m_{D}^{2}}{m_{c}},
\label{eq:Mel2}
\end{eqnarray}%
where $f_{D^{\ast }}$ and $f_{D}$ are their decay constants, whereas $%
\varepsilon _{\nu }$ is the polarization vector of the meson $D^{\ast +}$.
We model the vertex\ $\langle D^{0}(q)D^{\ast +}(p^{\prime },\varepsilon
)|M_{cc}^{+}(p,\epsilon )\rangle $ by the expression
\begin{eqnarray}
&&\langle D^{0}(q)D^{\ast +}(p^{\prime },\varepsilon )|M_{cc}^{+}(p,\epsilon
)\rangle =G(q^{2})\left[ (p\cdot p^{\prime })\right.  \notag \\
&&\left. \times (\varepsilon ^{\ast }\cdot \epsilon )-(p\cdot \varepsilon
^{\ast })(p^{\prime }\cdot \epsilon )\right],
\end{eqnarray}%
with $G(q^{2})$ being the strong coupling at the vertex $M_{cc}^{+}D^{0}D^{%
\ast +}$. Then, it is not difficult to show that
\begin{eqnarray}
&&\Pi _{\mu \nu }^{\mathrm{Phys}}(p,p^{\prime })=G(q^{2})\frac{fmf_{D^{\ast
}}m_{D^{\ast }}m_{D}^{2}f_{D}}{m_{c}(p^{2}-m^{2})(p^{\prime 2}-m_{D^{\ast
}}^{2})}  \notag \\
&&\times \frac{1}{(q^{2}-m_{D}^{2})}\left( \frac{m^{2}+m_{D^{\ast
}}^{2}-q^{2}}{2}g_{\mu \nu }-p_{\nu }p_{\mu }^{\prime }\right) +\cdots.
\notag \\
&&  \label{eq.Phys2}
\end{eqnarray}

The double Borel transformation of $\Pi _{\mu \nu }^{\mathrm{Phys}%
}(p,p^{\prime })$ over variables $p^{2}$ and $p^{\prime 2}$ yields
\begin{eqnarray}
&&\mathcal{B}\Pi _{\mu \nu }^{\mathrm{Phys}}(p,p^{\prime })=G(q^{2})\frac{%
fmf_{D^{\ast }}m_{D^{\ast }}m_{D}^{2}f_{D}}{m_{c}(q^{2}-m_{D}^{2})}%
e^{-m^{2}/M_{1}^{2}}  \notag \\
&&\times e^{-m_{D^{\ast }}^{2}/M_{2}^{2}}\left( \frac{m^{2}+m_{D^{\ast
}}^{2}-q^{2}}{2}g_{\mu \nu }-p_{\nu }p_{\mu }^{\prime }\right) +\cdots.
\notag \\
&&
\end{eqnarray}%
The function $\mathcal{B}\Pi _{\mu \nu }^{\mathrm{Phys}}(p,p^{\prime })$ is
the sum of two terms $\sim g_{\mu \nu }$ and $\sim p_{\nu }p_{\mu }^{\prime
} $, which may be utilized to obtain the required sum rule. For further
studies, we choose the invariant amplitude $\Pi ^{\mathrm{Phys}%
}(p^{2},p^{\prime 2},q^{2})$ corresponding to the structure $\sim g_{\mu \nu
}$. The Borel transformation of this amplitude constitutes the physical side
of the sum rule.

To determine the QCD side of the three-point sum rule, one should compute $%
\Pi _{\mu \nu }(p,p^{\prime })$ using quark propagators. As a result, one
gets
\begin{eqnarray}
&&\Pi _{\mu \nu }^{\mathrm{OPE}}(p,p^{\prime })=i^{2}\int
d^{4}xd^{4}ye^{i(p^{\prime }y-px)}\left\{ \mathrm{Tr}\left[ \gamma _{\nu
}S_{c}^{ia}(y-x)\right. \right.  \notag \\
&&\left. \times \gamma _{\mu }S_{d}^{ai}(x-y)\right] \mathrm{Tr}\left[
\gamma _{5}S_{u}^{bj}(x)\gamma _{5}S_{c}^{jb}(-x)\right]  \notag \\
&&\left. -\mathrm{Tr}\left[ \gamma _{\nu }S_{c}^{ib}(y-x)\gamma
_{5}S_{u}^{bj}(x)\gamma _{5}S_{c}^{ja}(-x)\gamma _{\mu }S_{d}^{ai}(x-y)%
\right] \right\} .  \notag \\
&&  \label{eq:CF4}
\end{eqnarray}

The correlation function $\Pi _{\mu \nu }^{\mathrm{OPE}}(p,p^{\prime })$ is
computed by taking into account terms up to dimension $6$, and has
structures identical to ones from $\Pi _{\mu \nu }^{\mathrm{Phys}%
}(p,p^{\prime })$. The explicit expression of $\Pi _{\mu \nu }^{\mathrm{OPE}%
}(p,p^{\prime })$ is rather lengthy, therefore we do not provide it here.

The double Borel transform of the invariant amplitude $\Pi ^{\mathrm{OPE}%
}(p^{2},p^{\prime 2},q^{2})$ which corresponds to the term $\sim g_{\mu \nu
} $ forms the QCD side of the sum rule. By equating the Borel transforms of
the amplitudes $\Pi ^{\mathrm{OPE}}(p^{2},p^{\prime 2},q^{2})$ and $\Pi ^{%
\mathrm{Phys}}(p^{2},p^{\prime 2},q^{2})$, and carrying out the continuum
subtraction, one gets the sum rule for the coupling $G(q^{2})$.

The amplitude $\Pi ^{\mathrm{OPE}}(p^{2},p^{\prime 2},q^{2})$ after the
Borel transformation and subtraction can be expressed in terms of the
spectral density $\rho (s,s^{\prime },q^{2})$ which is determined as a
relevant imaginary part of $\Pi _{\mu \nu }^{\mathrm{OPE}}(p,p^{\prime })$,
\begin{eqnarray}
&&\Pi (\mathbf{M}^{2},\mathbf{s}_{0},q^{2})=\int_{4m_{c}^{2}}^{s_{0}}ds%
\int_{m_{c}^{2}}^{s_{0}^{\prime }}ds^{\prime }\rho (s,s^{\prime },q^{2})
\notag \\
&&\times e^{-s/M_{1}^{2}}e^{-s^{\prime }/M_{2}^{2}}.  \label{eq:SCoupl}
\end{eqnarray}%
Here, $\mathbf{M}^{2}=(M_{1}^{2},M_{2}^{2})$ and $\mathbf{s}%
_{0}=(s_{0},s_{0}^{\prime })$ are the Borel and continuum threshold
parameters, respectively. The sum rule for $G(q^{2})$ is given by the
following expression
\begin{eqnarray}
&&G(q^{2})=\frac{2m_{c}}{fmf_{D^{\ast }}m_{D^{\ast }}m_{D}^{2}f_{D}}\frac{%
q^{2}-m_{D}^{2}}{m^{2}+m_{D^{\ast }}^{2}-q^{2}}  \notag \\
&&\times e^{m^{2}/M_{1}^{2}}e^{m_{D^{\ast }}^{2}/M_{2}^{2}}\Pi (\mathbf{M}%
^{2},\mathbf{s}_{0},q^{2}).  \label{eq:SRCoup}
\end{eqnarray}%
The coupling $G(q^{2})$ is a function of $q^{2}$ and parameters $(\mathbf{M}%
^{2},\mathbf{s}_{0})$: the latter, for simplicity, are not written down in
Eq.\ (\ref{eq:SRCoup}) as its arguments. In what follows, we use a new
variable $Q^{2}=-q^{2}$ and fix the obtained function by the notation $%
G(Q^{2})$.

The sum rule Eq.\ (\ref{eq:SRCoup}) depends on the mass and coupling of the
hadronic molecule $M_{cc}^{+}$, which are original results of the current
work and have been presented in Eq.\ (\ref{eq:Result1}). The equation (\ref%
{eq:SRCoup}) also contains the masses and decay constants of the mesons $%
D^{\ast +}$ and $D^{0}$. The masses of these mesons have been written down
in Eq.\ (\ref{eq:Masses}), whereas for their decay constants, we employ%
\begin{eqnarray}
f_{D^{\ast }} &=&(223.5\pm 8.4)~\mathrm{MeV},  \notag \\
f_{D} &=&(212.6\pm 0.7)~\mathrm{MeV}.  \label{eq:DecayCons}
\end{eqnarray}

Besides these parameters, for computation of $G(Q^{2})$ one should choose
working windows for $\mathbf{M}^{2}$ and $\mathbf{s}_{0}$ as well. The
restrictions used in such analysis are standard ones for sum rule
computations and have been considered above.
\begin{figure}[h]
\includegraphics[width=8.5cm]{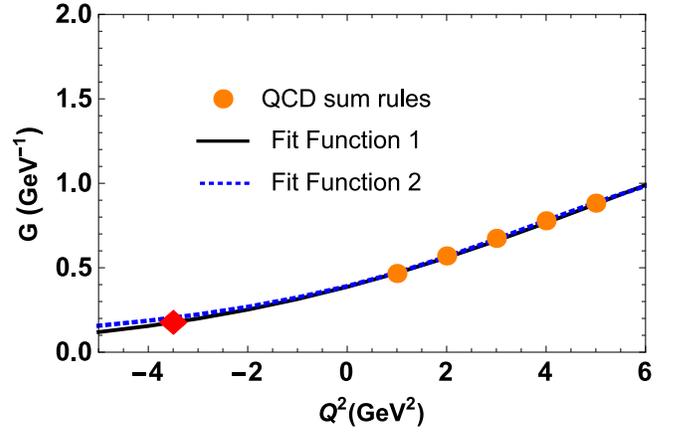}
\caption{The sum rule results and functions $F(Q^2)$ (FF 1) and $\overline{F}
(Q^2)$ (FF 2) for the strong coupling $G(Q^2)$. The red diamond fixes the
point $Q^2=-m_{D}^2$. }
\label{fig:Fit}
\end{figure}
The windows for $M_{1}^{2}$ and $s_{0}$ correspond to the $M_{cc}^{+}$
channel and are given by Eq.\ (\ref{eq:Regions}). The parameters $%
(M_{2}^{2},s_{0}^{\prime })$ for the $D^{\ast +}$ meson's channel vary
inside the intervals
\begin{equation}
M_{2}^{2}\in \lbrack 2,4]~\mathrm{GeV}^{2},\ s_{0}^{\prime }\in \lbrack
5.5,6.5]~\mathrm{GeV}^{2}.  \label{eq:Wind3}
\end{equation}

We calculate $G(Q^{2})$ at fixed $Q^{2}=1-6~\mathrm{GeV}^{2}$ and plot
obtained results in Fig.\ \ref{fig:Fit}. It is worth noting that at each $%
Q^{2}$ computations satisfy constraints imposed on parameters $\mathbf{M}%
^{2} $ and $\mathbf{s}_{0}$ by the sum rule analysis. Thus, in Fig.\ \ref%
{fig:StrongC} the coupling $G(Q^{2})$ is depicted as a function of the
parameters $M_{1}^{2}$ and $M_{2}^{2}$ at $Q^{2}=1~\mathrm{GeV}^{2}$ and
middle of the regions $s_{0}$ and $s_{0}^{\prime }$. A relative stability of
$G(1\ \mathrm{GeV}^{2})$ upon changing of $\mathbf{M}^{2}$ is evident:
Variations of $M_{1}^{2}$ and $M_{2}^{2}$ within explored regions do not
exceed $30\%$ of the central value for $G(1\ \mathrm{GeV}^{2})$.
Numerically, we find
\begin{equation}
G(1\ \mathrm{GeV}^{2})=0.48_{-0.09}^{+0.14}~\mathrm{GeV}^{-1}.
\end{equation}

\begin{figure}[h]
\includegraphics[width=8.8cm]{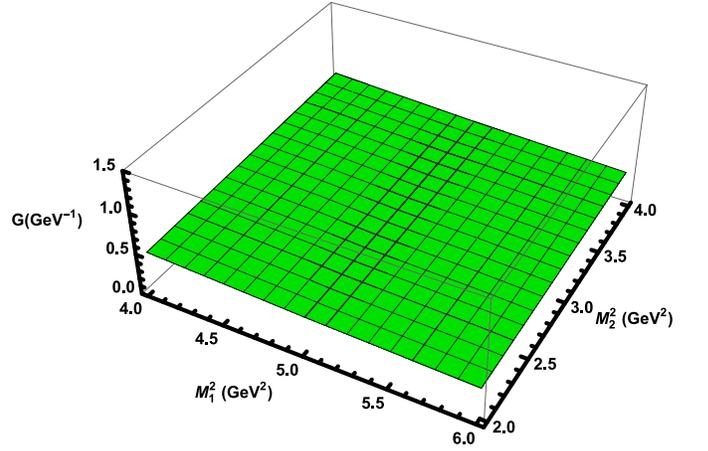}
\caption{ The strong coupling $G=G(1~\mathrm{GeV}^{2})$ as a function of the
Borel parameters $M_{1}^{2}$ and $M_{2}^{2}$ at $s_{0}=20.5~\mathrm{GeV}^{2}$
and $s_{0}^{\prime }=6~\mathrm{GeV}^{2}$.}
\label{fig:StrongC}
\end{figure}

The width of the process $M_{cc}^{+}\rightarrow D^{0}D^{\ast +}$ is
determined by the coupling $G$ at the mass shell $q^{2}=m_{D}^{2}$ of the
meson $D^{0}$, which cannot be calculated directly using the sum rule
method. To avoid this difficulty, we introduce fit functions $F(Q^{2})$ and $%
\overline{F}(Q^{2})$ that for the momenta $Q^{2}>0$ give results identical
to QCD sum rule's ones, but can be extrapolated to the region of $Q^{2}<0$
to fix $G$. We employ the fit functions $F(Q^{2})$ and $\overline{F}(Q^{2})$
given by the expressions%
\begin{equation}
F(Q^{2})=F_{0}\mathrm{\exp }\left[ c_{1}\frac{Q^{2}}{m^{2}}+c_{2}\left(
\frac{Q^{2}}{m^{2}}\right) ^{2}\right] ,  \label{eq:FitF}
\end{equation}%
and
\begin{equation}
\overline{F}(Q^{2})=\frac{\overline{F}_{0}}{\left( 1-\frac{Q^{2}}{m^{2}}%
\right) \left( 1-\sigma _{1}\frac{Q^{2}}{m^{2}}+\sigma _{2}\left( \frac{Q^{2}%
}{m^{2}}\right) ^{2}\right) },  \label{eq:FitFTilde}
\end{equation}%
where $F_{0}$, $c_{1}$ and $c_{2}$ and $\overline{F}_{0}$, $\sigma _{1}$ and
$\sigma _{2}$ are fitting parameters. From numerical computations, it is not
difficult to find that $F_{0}=0.39~\mathrm{GeV}^{-1},$ $c_{1}=3.29$ and $%
c_{2}=-1.95$. Similar analysis gives $\overline{F}_{0}=0.39~\mathrm{GeV}%
^{-1},$ $\sigma _{1}=2.11$ and $\sigma _{2}=2.97$. In Fig.\ \ref{fig:Fit},
along with the sum rule results for $G(Q^{2})$, we plot also the functions $%
F(Q^{2})$ and $\overline{F}(Q^{2})$. It is seen, that there are nice
agreements between the fit functions and QCD data. Their predictions for the
coupling $G$ are also very close to each other, and generate only small
additional uncertainties $\pm 0.01$ in $G$.

The functions $F(Q^{2})$ and $\overline{F}(Q^{2})$ at the $D^{0}$ meson's
mass shell lead to the average result
\begin{equation}
G=(0.192\pm 0.061)~\mathrm{GeV}^{-1},  \label{eq:Coupl1}
\end{equation}%
where the theoretical errors are the sum (in quadrature) of uncertainties
coming from sum rule computations $\pm 0.06$ and ones due to fitting
procedures. The width of decay $M_{cc}^{+}\rightarrow D^{0}D^{\ast +}$ is
determined by the expression
\begin{eqnarray}
&&\Gamma \left[ M_{cc}^{+}\rightarrow D^{0}D^{\ast +}\right] =G^{2}\frac{%
m_{D^{\ast }}^{2}\lambda \left( m,m_{D^{\ast }},m_{D}\right) }{24\pi }
\notag \\
&&\times \left( 3+\frac{2\lambda ^{2}\left( m,m_{D^{\ast }},m_{D}\right) }{%
m_{D^{\ast }}^{2}}\right) ,  \label{eq:DW1}
\end{eqnarray}%
where%
\begin{equation}
\lambda \left( a,b,c\right) =\frac{1}{2a}\sqrt{a^{4}+b^{4}+c^{4}-2\left(
a^{2}b^{2}+a^{2}c^{2}+b^{2}c^{2}\right) }.
\end{equation}%
Using the strong coupling from Eq.\ (\ref{eq:Coupl1}), one can evaluate
width of the process $M_{cc}^{+}\rightarrow D^{0}D^{\ast +}$%
\begin{equation}
\Gamma \left[ M_{cc}^{+}\rightarrow D^{0}D^{\ast +}\right] =(3.8\pm 1.7)~%
\mathrm{MeV}.  \label{eq:DW1Numeric}
\end{equation}

There are also other decay modes of the molecule $M_{cc}^{+}$, which produce
mesons $D^{0}D^{0}\pi ^{+}$ or $D^{0}D^{+}\pi ^{0}$. They run through
creation of intermediate scalar tetraquarks followed by their decays to a
pair of conventional mesons \cite{Agaev:2021vur}. These modes establish the
main mechanism for strong decays of $T_{cc}^{+}$, but are subdominant
processes for the molecule $M_{cc}^{+}$, and therefore can be neglected. Our
prediction for the width of $M_{cc}^{+}$ demonstrates that it is a
relatively wide resonance.


\section{Discussion and conclusions}

\label{sec:Conclusion}
We have calculated the mass and width of the doubly charmed axial-vector
state with quark content $cc\overline{u}\overline{d}$ by modeling it as the
hadronic molecule $M_{cc}^{+}=D^{0}D^{\ast +}$. Predictions obtained for the
mass $m=(4060\pm 130)~\mathrm{MeV}$ and width $\Gamma =(3.8\pm 1.7)~\mathrm{%
MeV}$ of this molecule exceed the LHCb data \cite{Aaij:2021vvq,LHCb:2021auc}%
. In our previous work, we carried out similar analysis by treating the
axial-vector $T_{cc}^{+}=cc\overline{u}\overline{d}$ state in the context of
the tetraquark model \cite{Agaev:2021vur}. Parameters of the exotic meson $%
T_{cc}^{+}$ agree nicely with data of the LHCb collaboration. Comparing with
each another results extracted from the sum rules in tetraquark and molecule
models, we see that the molecule $M_{cc}^{+}$ is heavier and wider than the
tetraquark structure.

Actually one might expect such outcome, because colored diquark and
antidiquark compact to form tightly-bound state, whereas interaction of two
colorless mesons is less intensive. Decays of tetraquarks and molecules to
conventional mesons also differ from each other. Indeed, in the case of a
tetraquark these processes require reorganization of its quark structure.
Contrary, a hadronic molecule's dissociation is free of such obstacles.
Hence, hadronic molecules are usually heavier and wider than their
tetraquark counterparts.

In this regards, it is instructive to recall a situation with the resonance $%
X_{0}(2900)$ discovered also by the LHCb collaboration. In Ref.\ \cite%
{Agaev:2020nrc}, we investigated $X_{0}$ and evaluated its parameters.
Results obtained for the mass and width of $X_{0}$ allowed us to interpret
it as the hadronic molecule $\overline{D}^{\ast }K^{\ast }$. We argued
additionally that a ground-state $1S$ scalar tetraquark with the same
content should have considerably smaller mass. This conclusion was supported
by analysis of Ref.\ \cite{He:2020jna}, in which $X_{0}$ was considered as
the radially excited $2S$ tetraquark $[ud][\overline{s}\overline{c}]$. The
mass difference between $1S$ and $2S$ particles equals there to $\approx 500~%
\mathrm{MeV}$. Because the molecule $\overline{D}^{\ast }K^{\ast }$ and $2S$
tetraquark $[ud][\overline{s}\overline{c}]$ have approximately equal masses,
the same estimate is valid for a mass gap between the molecule $\overline{D}%
^{\ast }K^{\ast }$ and ground state tetraquark. In the case of the exotic
mesons $M_{cc}^{+}$ and $T_{cc}^{+}$ this mass difference amounts to
approximately $200$ $\mathrm{MeV}$ being in a qualitative agreement with the
above analysis.

The molecule $M_{cc}^{+}$ was considered using the QCD sum rule method also
in other articles. Thus, in Ref.\ \cite{Dias:2011mi} the mass of $M_{cc}^{+}$
was estimated indirectly using the spectral sum rule prediction for the
ratio between the masses of $M_{cc}^{+}$ and resonance $X(3872)$. Let us
note that relevant calculations were carried by taking into account
condensates up to dimension-$6$. The prediction $m=(3872.2\pm 39.5)~\mathrm{%
MeV}$ obtained there for the mass of $M_{cc}^{+}$ is below two-meson
threshold and close to the LHCb data.

Direct sum rule computations of parameters of doubly charmed axial-vector
states $cc\overline{u}\overline{d}$ were performed in Refs.\ \cite%
{Xin:2021wcr,Wang:2017dtg}. In the tetraquark model the mass of such state
was predicted within the range \cite{Wang:2017dtg}
\begin{equation}
\widetilde{m}=(3900\pm 90)~\mathrm{MeV}.  \label{eq:Mass1}
\end{equation}%
This is higher than the LHCb data, and exceeds also our result $(3868\pm
124)~\mathrm{MeV}$ for this model from Ref.\ \cite{Agaev:2021vur}.

The axial-vector isoscalar and isovector molecules built of mesons $D^{0}$
and $D^{\ast +}$ were explored in Ref.\ \cite{Xin:2021wcr}, in which their
masses were found equal to%
\begin{eqnarray}
m_{\mathrm{I=0}} &=&(3880\pm 110)~\mathrm{MeV},  \notag \\
m_{\mathrm{I=1}} &=&(3890\pm 110)~\mathrm{MeV},  \label{eq:Masses2}
\end{eqnarray}%
respectively. The isoscalar molecule was interpreted as the LHCb resonance $%
T_{cc}^{+}$, or its essential component. Of course, comparing Eqs.\ (\ref%
{eq:Result1}) and (\ref{eq:Masses2}) one sees overlapping regions for the
mass of $M_{cc}^{+}$, but there are essential differences between relevant
central values.

It is interesting that masses of the tetraquark and molecule states from
Eqs.\ (\ref{eq:Mass1}) and (\ref{eq:Masses2}) coincide with each other. This
fact may be explained by different choices for the
renormalization/factorization scale $\mu $ used to evolve vacuum condensates
and quark masses. Indeed, $\widetilde{m}$ was obtained at $\mu =1.3~\mathrm{%
GeV}$, whereas $m_{\mathrm{I}}$ extracted by employing the scale $\mu =1.4~%
\mathrm{GeV}$. Different scales presumably eliminate a typical mass gap
between tetraquark and molecule structures. The choice for the scale $\mu
>1\ \mathrm{GeV}$ may also generate the discrepancy between Eq.\ (\ref%
{eq:Masses2}) and our result for the molecule $D^{0}D^{\ast +}$. It is worth
to note that Eq.\ (\ref{eq:Result1}) have been obtained at $\mu =1~\mathrm{%
GeV}$, which is necessary for leading-order QCD calculations.

To fix the scale $\mu $ unambiguously and damp sensitivity of results
against its variations, one needs to find physical quantities with the
next-to-leading order (NLO) accuracy: The NLO results have enhanced
predictive power, and their comparisons with data lead to more reliable
conclusions. The factorized NLO perturbative corrections to doubly heavy
exotic mesons' masses and couplings were computed in Ref.\ \cite%
{Albuquerque:2022weq} in QCD Inverse Laplace sum rule approach. These
corrections are important to legitimate the choice of heavy quark masses
used in relevant studies, though in the $\overline{\mathrm{MS}}$ scheme NLO
effects themselves are small \cite{Albuquerque:2022weq}. The authors
explained by this fact success of corresponding leading-order QCD analyses.
It is possible to carry out similar NLO computations in the context of the
two-point sum rule method to remove ambiguities in the choice of the scale $%
\mu $ while calculating parameters of $T_{cc}^{+}$ and $M_{cc}^{+}$, which,
however, are beyond the scope of the current article.

Summing up, investigations of $T_{cc}^{+}$ and $M_{cc}^{+}$ in the framework
of QCD sum rule method lead for these states to wide diversity of
predictions. The sum rule method relies on fundamental principles of QCD and
uses universal vacuum condensates to extract parameters of various hadrons,
nevertheless it suffers from theoretical errors which make difficult
unambiguous interpretation of obtained results. Thus, predictions for the
mass and width of the hadronic molecule $M_{cc}^{+}$ obtained in the current
article differ from relevant LHCb data, but these differences remain within $1.5
$ and $2$ standard deviations, respectively. Therefore, though our studies demonstrate
that a preferable assignment for the LHCb resonance is the tetraquark model $%
T_{cc}^{+}$, within the theoretical uncertainties, they also do not rule out the  molecule picture $%
M_{cc}^{+}\equiv D^{0}D^{\ast +}$.

Controversial predictions for parameters of the molecule $M_{cc}^{+}$ were
made in the context of alternative methods \cite%
{Meng:2021jnw,Ling:2021bir,Chen:2021vhg} as well. Results for the full width
of the $M_{cc}^{+}$ obtained in papers \cite{Meng:2021jnw,Ling:2021bir} are
rather small compared with the LHCb data. At the same time, a nice agreement
with recent measurements was declared in Ref.\ \cite{Chen:2021vhg}.
Moreover, in this work the authors predicted existence of another doubly
charmed resonance with the mass $m=3876~\mathrm{MeV}$ and width $\Gamma =412~%
\mathrm{keV}$.

As is seen, even in the context of same models and methods, theoretical
investigations sometimes lead to contradictory predictions for the
parameters of the molecule $M_{cc}^{+}$: New efforts are required to settle
existing problems. Additionally, more accurate LHCb data are necessary for
the full width of the doubly charmed state $T_{cc}^{+}$ to compare with
different theoretical results.

\begin{widetext}

\appendix*

\section{ The propagators $S_{q(Q)}(x)$ and invariant amplitude $\Pi
(M^{2},s_{0})$}

\renewcommand{\theequation}{\Alph{section}.\arabic{equation}} \label{sec:App}

In the current article, for the light quark propagator $S_{q}^{ab}(x)$, we
employ the following expression
\begin{eqnarray}
&&S_{q}^{ab}(x)=i\delta _{ab}\frac{\slashed x}{2\pi ^{2}x^{4}}-\delta _{ab}%
\frac{m_{q}}{4\pi ^{2}x^{2}}-\delta _{ab}\frac{\langle \overline{q}q\rangle
}{12}+i\delta _{ab}\frac{\slashed xm_{q}\langle \overline{q}q\rangle }{48}%
-\delta _{ab}\frac{x^{2}}{192}\langle \overline{q}g_{s}\sigma Gq\rangle
\notag \\
&&+i\delta _{ab}\frac{x^{2}\slashed xm_{q}}{1152}\langle \overline{q}%
g_{s}\sigma Gq\rangle -i\frac{g_{s}G_{ab}^{\alpha \beta }}{32\pi ^{2}x^{2}}%
\left[ \slashed x{\sigma _{\alpha \beta }+\sigma _{\alpha \beta }}\slashed x%
\right] -i\delta _{ab}\frac{x^{2}\slashed xg_{s}^{2}\langle \overline{q}%
q\rangle ^{2}}{7776}  \notag \\
&&-\delta _{ab}\frac{x^{4}\langle \overline{q}q\rangle \langle
g_{s}^{2}G^{2}\rangle }{27648}+\cdots .
\end{eqnarray}%
For the heavy quark $Q=c$, we use the propagator $S_{Q}^{ab}(x)$
\begin{eqnarray}
&&S_{Q}^{ab}(x)=i\int \frac{d^{4}k}{(2\pi )^{4}}e^{-ikx}\Bigg \{\frac{\delta
_{ab}\left( {\slashed k}+m_{Q}\right) }{k^{2}-m_{Q}^{2}}-\frac{%
g_{s}G_{ab}^{\alpha \beta }}{4}\frac{\sigma _{\alpha \beta }\left( {\slashed %
k}+m_{Q}\right) +\left( {\slashed k}+m_{Q}\right) \sigma _{\alpha \beta }}{%
(k^{2}-m_{Q}^{2})^{2}}  \notag \\
&&+\frac{g_{s}^{2}G^{2}}{12}\delta _{ab}m_{Q}\frac{k^{2}+m_{Q}{\slashed k}}{%
(k^{2}-m_{Q}^{2})^{4}}+\frac{g_{s}^{3}G^{3}}{48}\delta _{ab}\frac{\left( {%
\slashed k}+m_{Q}\right) }{(k^{2}-m_{Q}^{2})^{6}}\left[ {\slashed k}\left(
k^{2}-3m_{Q}^{2}\right) +2m_{Q}\left( 2k^{2}-m_{Q}^{2}\right) \right] \left(
{\slashed k}+m_{Q}\right) +\cdots \Bigg \}.  \notag \\
&&
\end{eqnarray}

Here, we have used the short-hand notations
\begin{equation}
G_{ab}^{\alpha \beta }\equiv G_{A}^{\alpha \beta }\lambda _{ab}^{A}/2,\ \
G^{2}=G_{\alpha \beta }^{A}G_{A}^{\alpha \beta },\ G^{3}=f^{ABC}G_{\alpha
\beta }^{A}G^{B\beta \delta }G_{\delta }^{C\alpha },
\end{equation}%
where $G_{A}^{\alpha \beta }$ is the gluon field strength tensor, $\lambda
^{A}$ and $f^{ABC}$ are the Gell-Mann matrices and structure constants of
the color group $SU_{c}(3)$, respectively. The indices $A,B,C$ run in the
range $1,2,\ldots 8$.

The invariant amplitude $\Pi (M^{2},s_{0})$ obtained after the Borel
transformation and subtraction procedures is given by Eq.\ (\ref{eq:InvAmp})%
\begin{equation*}
\Pi (M^{2},s_{0})=\int_{4m_{c}^{2}}^{s_{0}}ds\rho ^{\mathrm{OPE}%
}(s)e^{-s/M^{2}}+\Pi (M^{2}),
\end{equation*}%
where the spectral density $\rho ^{\mathrm{OPE}}(s)$ and the function $\Pi
(M^{2})$ are determined by formulas
\begin{equation}
\rho ^{\mathrm{OPE}}(s)=\rho ^{\mathrm{pert.}}(s)+\sum_{N=3}^{8}\rho ^{%
\mathrm{DimN}}(s),\ \ \Pi (M^{2})=\sum_{N=6}^{10}\Pi ^{\mathrm{DimN}}(M^{2}),
\label{eq:A1}
\end{equation}%
respectively. The components of $\rho ^{\mathrm{OPE}}(s)$ and $\Pi (M^{2})$
are given by the expressions%
\begin{equation}
\rho ^{\mathrm{DimN}}(s)=\int_{0}^{1}d\alpha \int_{0}^{1-a}d\beta \rho ^{%
\mathrm{DimN}}(s,\alpha ,\beta ),\ \ \rho ^{\mathrm{DimN}}(s)=\int_{0}^{1}d%
\alpha \rho ^{\mathrm{DimN}}(s,\alpha ),  \label{eq:A2}
\end{equation}%
and
\begin{equation}
\Pi ^{\mathrm{DimN}}(M^{2})=\int_{0}^{1}d\alpha \int_{0}^{1-a}d\beta \Pi ^{%
\mathrm{DimN}}(M^{2},\alpha ,\beta ),\ \ \Pi ^{\mathrm{DimN}%
}(M^{2})=\int_{0}^{1}d\alpha \Pi ^{\mathrm{DimN}}(M^{2},\alpha ).
\label{eq:A4}
\end{equation}%
depending on whether $\rho $ and $\Pi (M^{2})$ are functions of $\alpha $
and $\beta $ or only of $\alpha $. In Eqs.\ (\ref{eq:A2}) and (\ref{eq:A4})
variables $\alpha $ and $\beta $ are Feynman parameters.

The perturbative and nonperturbative components of the spectral density $%
\rho ^{\mathrm{pert.}}(s,\alpha ,\beta )$ and $\rho ^{\mathrm{Dim3(4,5,6,7,8)%
}}(s,\alpha ,\beta )$ \ have the forms:
\begin{eqnarray}
&&\rho ^{\mathrm{pert.}}(s,\alpha ,\beta )=\frac{\Theta (L_{1})}{49152\pi
^{6}L^{2}N_{1}^{8}}\left[ m_{c}^{2}N_{2}-s\alpha \beta L\right] ^{2}\left\{
s^{2}\alpha ^{2}\beta ^{2}L^{3}\left[ 18\beta ^{2}+18\alpha (\alpha
-1)-\beta (18+325\alpha )\right] +m_{c}^{4}N_{1}^{2}\left[ 18\beta
^{5}\right. \right.  \notag \\
&&\left. +18\alpha ^{3}(\alpha -1)^{2}+\beta ^{4}(37\alpha -36)+\beta
^{2}\alpha (54-110\alpha +29\alpha ^{2})+\beta ^{3}(18-91\alpha +29\alpha
^{2})+\beta \alpha ^{2}(54-91\alpha +37\alpha ^{2})\right]  \notag \\
&&-2m_{c}^{2}s\alpha \beta \left[ 18\beta ^{7}+18\alpha ^{3}(\alpha
-1)^{4}-3\beta \alpha ^{2}(\alpha -1)^{3}(18+13\alpha )-3\beta
^{6}(24+13\alpha )+\beta ^{5}(108+63\alpha -296\alpha ^{2})\right.  \notag \\
&&-2\beta ^{2}\alpha (\alpha -1)^{2}(-27-3\alpha +148\alpha ^{2})+\beta
^{4}(-72+45\alpha +598\alpha ^{2}-571\alpha ^{3})+\beta ^{3}\left(
18-123\alpha -254\alpha ^{2}+930\alpha ^{3}\right.  \notag \\
&&\left. \left. \left. -517\alpha ^{4}\right) \right] \right\} ,
\end{eqnarray}%
\begin{equation}
\rho ^{\mathrm{Dim3}}(s,\alpha ,\beta )=\frac{m_{c}\Theta (L_{1})}{512\pi
^{4}N_{1}^{5}}\left[ m_{c}^{2}N_{2}-s\alpha \beta L\right] \left\{ 2\langle
\overline{d}d\rangle (11\beta -\alpha )\left[ m_{c}^{2}N_{2}-2s\alpha \beta L%
\right] -\langle \overline{u}u\rangle (\beta -11\alpha )\left[
m_{c}^{2}N_{2}-3s\alpha \beta L\right] \right\} ,
\end{equation}%
\begin{eqnarray}
&&\rho ^{\mathrm{Dim4}}(s,\alpha ,\beta )=\frac{\langle \alpha _{s}G^{2}/\pi
\rangle \Theta (L_{1})}{73728\pi ^{4}(\beta -1)L^{2}N_{1}^{6}}\left\{
-s^{2}(\beta -1)\alpha ^{2}\beta ^{2}L^{2}\left[ 18\beta ^{4}+6\alpha
(\alpha -1)^{2}(3\alpha -2)+3\beta ^{3}(417\alpha -16)\right. \right.  \notag
\\
&&\left. +\beta ^{2}(42-1521\alpha -91\alpha ^{2})-3\beta (4-94\alpha
-333\alpha ^{2}+423\alpha ^{3})\right] -m_{c}^{4}N_{1}^{2}\left[ 18\beta
^{8}+\beta ^{7}(611\alpha -108)\right.  \notag \\
&&+\beta ^{6}(228-2053\alpha +1918\alpha ^{2})-6\alpha ^{3}(\alpha
-1)^{2}(2-9\alpha +3\alpha ^{2}+2\alpha ^{3})+\beta ^{5}(-216+2567\alpha
-5375\alpha ^{2}+2181\alpha ^{3})  \notag \\
&&+\beta ^{4}(90-1455\alpha +5518\alpha ^{2}-5083\alpha ^{3}+857\alpha
^{4})+\beta \alpha ^{2}(-36+342\alpha -555\alpha ^{2}+248\alpha
^{3}+23\alpha ^{4}-22\alpha ^{5})  \notag \\
&&-\beta ^{3}(12-366\alpha +2565\alpha ^{2}-4315\alpha ^{3}+1686\alpha
^{4}+94\alpha ^{5})-\beta ^{2}\alpha \left( 36-540\alpha +1743\alpha
^{2}-1306\alpha ^{3}+16\alpha ^{4}\right.  \notag \\
&&\left. +105\alpha ^{5}\right) +2m_{c}^{2}s\alpha \beta \left[ 15\beta
^{10}+13\beta ^{9}(74\alpha -9)+\beta ^{8}(372-5060\alpha +3754\alpha
^{2})-3\alpha ^{3}(\alpha -1)^{4}(4-14\alpha +5\alpha ^{2}+2\alpha
^{3})\right.  \notag \\
&&+\beta ^{7}\left( -630+10956\alpha -17152\alpha ^{2}+6047\alpha
^{3}\right) -\beta \alpha ^{2}(\alpha -1)^{3}(-36+348\alpha +174\alpha
^{2}-439\alpha ^{3}+11\alpha ^{4})  \notag \\
&&+\beta ^{6}\left( 615-12500\alpha +31809\alpha ^{2}-23517\alpha
^{3}+4514\alpha ^{4}\right) +\beta ^{5}\left( -345+7970\alpha -30691\alpha
^{2}+36238\alpha ^{3}-13408\alpha ^{4}+160\alpha ^{5}\right)  \notag \\
&&-\beta ^{2}\alpha (\alpha -1)^{2}\left( 36-672\alpha +1809\alpha
^{2}+1220\alpha ^{3}-2788\alpha ^{4}+538\alpha ^{5}\right) +\beta ^{4}\left(
102-2772\alpha +16465\alpha ^{2}-28426\alpha ^{3}+14658\alpha ^{4}\right.
\notag \\
&&\left. \left. \left. +2648\alpha ^{5}-2675\alpha ^{6}\right) -\beta
^{3}\left( 12-480\alpha +4893\alpha ^{2}-12403\alpha ^{3}+7946\alpha
^{4}+6091\alpha ^{5}-8153\alpha ^{6}+2094\alpha ^{7}\right) \right] \right\}
,
\end{eqnarray}%
\begin{eqnarray}
\rho ^{\mathrm{Dim5}}(s,\alpha ,\beta ) &=&-\frac{m_{c}\Theta (L_{1})L}{%
1024\pi ^{4}N_{1}^{4}}\left\{ -\langle \overline{u}g_{s}\sigma Gu\rangle
(\beta -11\alpha )\left[ m_{c}^{2}N_{2}-2s\alpha \beta L\right] \right.
\notag \\
&&\left. +\langle \overline{d}g_{s}\sigma Gd\rangle (11\beta -\alpha )\left[
m_{c}^{2}N_{2}-3s\alpha \beta L\right] \right\} ,
\end{eqnarray}%
\begin{eqnarray}
&&\rho _{1}^{\mathrm{Dim6}}(M^{2},\alpha ,\beta )=\frac{\langle
g_{s}^{3}G^{3}\rangle \Theta (L_{1})}{45\cdot 2^{19}\pi ^{6}L^{2}N_{1}^{7}}%
\left\{ 36m_{c}^{2}N_{1}^{2}\left[ 12\beta ^{9}-5\beta ^{4}\alpha
^{5}-4\beta ^{3}\alpha ^{5}(\alpha -1)+12\beta \alpha ^{5}(\alpha
-1)^{3}+12\alpha ^{6}(\alpha -1)^{3}\right. \right.  \notag \\
&&+4\beta ^{8}(8\alpha -9)+\beta ^{2}\alpha ^{5}(13-18\alpha +5\alpha
^{2})+\beta ^{7}(36-76\alpha +45\alpha ^{2})+3\beta ^{5}\alpha (-4+11\alpha
-12\alpha ^{2}+5\alpha ^{3})+2\beta ^{6}\left( -6\right.  \notag \\
&&\left. \left. +28\alpha -39\alpha ^{2}+18\alpha ^{3}\right) \right]
-s\alpha \beta L^{2}\left[ 216\beta ^{9}+216\alpha ^{6}(\alpha
-1)^{3}-24\beta \alpha ^{5}(\alpha -1)^{2}(3+73\alpha )-24\beta
^{8}(27+73\alpha )\right.  \notag \\
&&+\beta ^{2}\alpha ^{4}(\alpha -1)^{2}(144+2773\alpha )+\beta ^{3}\alpha
^{4}(-543-1759\alpha +2302\alpha ^{2})+\beta ^{7}(648+3432\alpha +2611\alpha
^{2})  \notag \\
&&-3\beta ^{4}\alpha ^{2}(-48+235\alpha -544\alpha ^{2}+357\alpha
^{3})+2\beta ^{6}(-108-804\alpha -2539\alpha ^{2}+989\alpha ^{3})  \notag \\
&&\left. -\beta ^{5}\alpha (72-2323\alpha +1273\alpha ^{2}+1233\alpha ^{3})
\right] +72m_{c}^{2}\left[ 2\beta ^{13}+2\alpha ^{8}(\alpha -1)^{5}-\beta
\alpha ^{7}(\alpha -1)^{4}(6+\alpha )-\beta ^{12}(10+\alpha )\right.  \notag
\\
&&+\beta ^{11}(20-2\alpha -5\alpha ^{2})-\beta ^{2}\alpha ^{6}(\alpha
-1)^{3}(-6+10\alpha +5\alpha ^{2})+\beta ^{3}\alpha ^{5}(\alpha
-1)^{3}(2-21\alpha +8\alpha ^{2})+\beta ^{4}\alpha ^{5}(\alpha -1)^{2}
\notag \\
&&\times (12-46\alpha +31\alpha ^{2})+\beta ^{10}(-20+18\alpha +5\alpha
^{2}+8\alpha ^{3})+\beta ^{5}\alpha ^{3}(\alpha -1)^{2}(-2+8\alpha -30\alpha
^{2}+41\alpha ^{3})+\beta ^{9}\left( 10-32\alpha \right.  \notag \\
&&\left. +21\alpha ^{2}-45\alpha ^{3}+31\alpha ^{4}\right) +\beta
^{8}(-2+23\alpha -43\alpha ^{2}+89\alpha ^{3}-108\alpha ^{4}+41\alpha
^{5})+\beta ^{7}\alpha \left( -6+28\alpha -77\alpha ^{2}+135\alpha
^{3}\right.  \notag \\
&&\left. \left. \left. -112\alpha ^{4}+32\alpha ^{5}\right) +\beta
^{6}\alpha ^{2}(-6+27\alpha -70\alpha ^{2}+109\alpha ^{3}-92\alpha
^{4}+32\alpha ^{5})\right] \right\} ,
\end{eqnarray}%
\begin{eqnarray}
&&\rho _{1}^{\mathrm{Dim7}}(M^{2},\alpha ,\beta )=\frac{m_{c}\langle \alpha
_{s}G^{2}/\pi \rangle \Theta (L_{1})}{4608\pi ^{2}N_{1}^{4}}\left\{ 2\langle
\overline{d}d\rangle \left[ 25\beta ^{5}-2\beta ^{4}(14+19\alpha )+\beta
^{3}(3+65\alpha -68\alpha ^{2})\right. \right.  \notag \\
&&\left. +\beta ^{2}\alpha (-27+84\alpha -56\alpha ^{2})+\beta \alpha
^{2}(-27+53\alpha -26\alpha ^{2})+\alpha ^{3}(3-4\alpha +\alpha ^{2})\right]
\notag \\
&&+\langle \overline{u}u\rangle \left[ -5\beta ^{5}+\beta ^{4}(8+22\alpha
)+\beta \alpha ^{2}(27-37\alpha +10\alpha ^{2})+\alpha ^{3}(-3-16\alpha
+19\alpha ^{2})\right.  \notag \\
&&\left. \left. +\beta ^{2}\alpha (27-72\alpha +34\alpha ^{2})+\beta
^{3}(-3-49\alpha +46\alpha ^{2})\right] \right\} ,
\end{eqnarray}%
\begin{equation}
\rho _{1}^{\mathrm{Dim8}}(M^{2},\alpha ,\beta )=-\frac{\langle \alpha
_{s}G^{2}/\pi \rangle ^{2}}{24576\pi ^{2}N_{1}^{3}}\Theta (L_{1})\alpha
\beta L.
\end{equation}%
The components $\rho ^{\mathrm{Dim6(7,8)}}(s,\alpha )$ are given by the
formulas%
\begin{equation}
\rho _{2}^{\mathrm{Dim6}}(s,\alpha )=\frac{\langle \overline{d}d\rangle
\langle \overline{u}u\rangle }{192\pi ^{2}}\Theta (L_{2})\left[
13m_{c}^{2}+3s\alpha (\alpha -1)\right] ,\ \
\end{equation}%
\begin{equation}
\rho _{2}^{\mathrm{Dim7}}(s,\alpha )=\frac{m_{c}\langle \alpha _{s}G^{2}/\pi
\rangle }{9216\pi ^{2}}\Theta (L_{2})\left[ \langle \overline{u}u\rangle
(1-12\alpha )+\langle \overline{d}d\rangle (-22+24\alpha )\right] ,\
\end{equation}%
and 
\begin{equation}
\rho _{2}^{\mathrm{Dim8}}(s,\alpha )=\frac{\langle \overline{d}d\rangle
\langle \overline{u}g_{s}\sigma Gu\rangle }{48\pi ^{2}}\Theta (L_{2})\alpha
(\alpha -1).\
\end{equation}

Components of the function $\Pi (M^{2})$ are:%
\begin{eqnarray}
&&\Pi ^{\mathrm{Dim6}}(M^{2},\alpha ,\beta )=\frac{m_{c}^{4}\langle
g_{s}^{3}G^{3}\rangle }{45\cdot 2^{18}\pi ^{6}\alpha \beta (\beta
-1)L^{3}N_{1}^{4}}\exp \left[ -\frac{m_{c}^{2}N_{2}}{M^{2}\alpha \beta L}%
\right] \left[ 29\beta ^{11}-(\alpha -1)^{3}\alpha ^{7}(99+70\alpha )\right.
\notag \\
&&+\beta ^{10}(-186+113\alpha )+\beta ^{9}(384-447\alpha -50\alpha
^{2})-\beta (\alpha -1)^{2}\alpha ^{6}(-108+212\alpha +111\alpha ^{2})
\notag \\
&&+\beta ^{8}(-326+663\alpha +200\alpha ^{2}-447\alpha ^{3})+\beta
^{6}\alpha (108+340\alpha -1886\alpha ^{2}+2485\alpha ^{3}-1042\alpha ^{4})
\notag \\
&&+\beta ^{7}(99-437\alpha -370\alpha ^{2}+1510\alpha ^{3}-880\alpha
^{4})+\beta ^{5}\alpha ^{2}(-120+1030\alpha -2607\alpha ^{2}+2577\alpha
^{3}-740\alpha ^{4})  \notag \\
&&+\beta ^{4}\alpha ^{3}(-207+1209\alpha -2358\alpha ^{2}+1593\alpha
^{3}-377\alpha ^{4})+\beta ^{3}\alpha ^{4}\left( -270+943\alpha -1143\alpha
^{2}\right.  \notag \\
&&\left. \left. +527\alpha ^{3}-120\alpha ^{4}\right) +\beta ^{2}\alpha
^{5}(-120+112\alpha +179\alpha ^{2}-74\alpha ^{3}-97\alpha ^{4})\right] ,
\end{eqnarray}%
\begin{eqnarray}
&&\Pi ^{\mathrm{Dim7}}(M^{2},\alpha ,\beta )=\frac{m_{c}\langle \alpha
_{s}G^{2}/\pi \rangle }{4608\pi ^{2}M^{2}\alpha \beta L^{2}N_{1}^{4}}\left\{
M^{2}\alpha ^{2}\beta ^{2}L^{3}[2\langle \overline{d}d\rangle (11\beta
^{2}-\alpha ^{2})+\langle \overline{u}u\rangle (11\alpha ^{2}-\beta
^{2})]\right.  \notag \\
&&+m_{c}^{4}\langle \overline{u}u\rangle N_{1}^{2}\exp \left[ -\frac{%
m_{c}^{2}N_{2}}{M^{2}\alpha \beta L}\right] \left[ 10\beta ^{5}+10\alpha
^{4}(\alpha -1)+\beta ^{4}(-10+19\alpha )+\beta ^{3}\alpha (-20+29\alpha
)\right.  \notag \\
&&\left. +\beta \alpha ^{3}(-20+31\alpha )+\beta ^{2}\alpha
^{2}(-20+41\alpha )\right] +m_{c}^{2}\langle \overline{d}d\rangle \exp \left[
-\frac{m_{c}^{2}N_{2}}{M^{2}\alpha \beta L}\right] \left[ m_{c}^{2}N_{1}^{2}%
\left( 10\beta ^{5}+10\alpha ^{4}(\alpha -1)\right. \right.  \notag \\
&&\left. +\beta \alpha ^{3}(-20+19\alpha )+\beta ^{2}\alpha
^{2}(-20+29\alpha )+\beta ^{4}(-10+31\alpha )+\beta ^{3}\alpha (-20+41\alpha
)\right)  \notag \\
&&-M^{2}\alpha \beta \left( 10\beta ^{7}+10\alpha ^{4}(\alpha -1)^{3}+\alpha
^{3}\beta (\alpha -1)^{2}(-20+29\alpha )+\beta ^{6}(-30+41\alpha )+2\beta
^{5}(15-51\alpha +41\alpha ^{2})\right.  \notag \\
&&+\beta ^{2}\alpha ^{2}(-20+99\alpha -137\alpha ^{2}+58\alpha ^{3})+\beta
^{3}\alpha (-20+111\alpha -180\alpha ^{2}+89\alpha ^{3})  \notag \\
&&\left. \left. \left. +\beta ^{4}(-10+81\alpha -173\alpha ^{2}+101\alpha
^{3})\right) \right] \right\} ,
\end{eqnarray}

\begin{eqnarray}
&&\Pi ^{\mathrm{Dim8}}(M^{2},\alpha ,\beta )=\frac{\langle \alpha
_{s}G^{2}/\pi \rangle ^{2}}{27\cdot 2^{13}\pi ^{2}M^{4}\alpha ^{2}\beta
^{2}L^{4}N_{1}^{3}}\left\{ 3M^{4}\alpha ^{3}\beta ^{3}L^{5}+m_{c}^{2}\exp %
\left[ -\frac{m_{c}^{2}N_{2}}{M^{2}\alpha \beta L}\right] \left\{
8m_{c}^{4}\alpha ^{2}\beta ^{2}N_{1}^{2}\right. \right.  \notag \\
&&\times \left[ 2\beta ^{3}+2\alpha ^{2}(\alpha -1)+\alpha \beta (-4+5\alpha
)+\beta ^{2}(-2+5\alpha )\right] -8m_{c}^{2}M^{2}\left[ 3\beta ^{11}+3\alpha
^{6}(\alpha -1)^{5}+3\beta ^{10}(-5+7\alpha )\right.  \notag \\
&&+3\beta \alpha ^{5}(\alpha -1)^{4}(-4+7\alpha )+\beta ^{9}(30-96\alpha
+74\alpha ^{2})+\beta ^{2}\alpha ^{4}(\alpha -1)^{3}(21-84\alpha +74\alpha
^{2})  \notag \\
&&+3\beta ^{8}(-10+58\alpha -102\alpha ^{2}+57\alpha ^{3})+\beta ^{3}\alpha
^{3}(\alpha -1)^{2}(-24+153\alpha -295\alpha ^{2}+171\alpha ^{3})  \notag \\
&&+\beta ^{4}\alpha ^{2}(\alpha -1)^{2}(-21+159\alpha -385\alpha
^{2}+286\alpha ^{3})+\beta ^{7}(15-156\alpha +495\alpha ^{2}-637\alpha
^{3}+286\alpha ^{4})  \notag \\
&&+\beta ^{5}\alpha (-12+147\alpha -625\alpha ^{2}+1215\alpha
^{3}-1091\alpha ^{4}+366\alpha ^{5})+\beta ^{6}\left( -3+69\alpha -389\alpha
^{2}+914\alpha ^{3}\right.  \notag \\
&&\left. \left. -957\alpha ^{4}+366\alpha ^{5}\right) \right] +M^{4}\alpha
\beta \left[ 48\beta ^{9}+48\alpha ^{4}(\alpha -1)^{5}+3\beta
^{8}(-80+71\alpha )+3\beta \alpha ^{3}(\alpha -1)^{4}(-23+71\alpha )\right.
\notag \\
&&+2\beta ^{2}\alpha ^{2}(\alpha -1)^{3}(21-180\alpha +239\alpha ^{2})+\beta
^{7}(480-921\alpha +478\alpha ^{2})+6\beta ^{6}(-80+259\alpha -299\alpha
^{2}+120\alpha ^{3})  \notag \\
&&+3\beta ^{3}\alpha (\alpha -1)^{2}(-23+116\alpha -317\alpha ^{2}+240\alpha
^{3})+3\beta ^{5}(80-422\alpha +852\alpha ^{2}-797\alpha ^{3}+287\alpha ^{4})
\notag \\
&&\left. \left. +\beta ^{4}(-48+489\alpha -1684\alpha ^{2}+2970\alpha
^{3}-2588\alpha ^{4}+861\alpha ^{5})\right] \right\} .
\end{eqnarray}

The dimension $9$ contribution to the correlation function is equal to zero.
The $\mathrm{Dim10}$ term is exclusively of the type (\ref{eq:A4}) and has
two components $\Pi _{1}^{\mathrm{Dim10}}(M^{2},\alpha ,\beta )$ and $\Pi
_{2}^{\mathrm{Dim10}}(M^{2},\alpha )$
\begin{eqnarray}
&&\Pi _{1}^{\mathrm{Dim10}}(M^{2},\alpha ,\beta )=\frac{\langle \alpha
_{s}G^{2}/\pi \rangle \langle g_{s}^{3}G^{3}\rangle }{135\cdot 2^{16}\pi
^{2}M^{8}\alpha ^{4}\beta ^{4}(\beta -1)^{2}L^{6}N_{1}^{4}}\exp \left[ -%
\frac{m_{c}^{2}N_{2}}{M^{2}\alpha \beta L}\right] \left\{ 36M^{8}\alpha
^{3}\beta ^{3}L^{6}R_{1}(\alpha ,\beta )\right.  \notag \\
&&-m_{c}^{8}\alpha \beta (\beta -1)^{2}N_{1}^{4}R_{2}(\alpha ,\beta
)+m_{c}^{6}M^{2}(\beta -1)^{2}N_{1}^{3}R_{3}(\alpha ,\beta
)-2m_{c}^{4}M^{4}\alpha \beta (\beta -1)^{2}N_{1}^{2}R_{4}(\alpha ,\beta )
\notag \\
&&\left. -2m_{c}^{2}M^{6}\alpha ^{2}\beta ^{2}L^{2}R_{5}(\alpha ,\beta
)\right\} ,
\end{eqnarray}%
and
\begin{eqnarray}
&&\Pi _{2}^{\mathrm{Dim10}}(M^{2},\alpha )=\frac{m_{c}^{2}\langle \alpha
_{s}G^{2}/\pi \rangle \langle \overline{d}d\rangle \langle \overline{u}%
u\rangle }{432M^{4}\alpha ^{3}(\alpha -1)^{3}}\exp \left[ -\frac{m_{c}^{2}}{%
M^{2}\alpha (\alpha -1)}\right] \left[ m_{c}^{2}+2M^{2}\alpha (\alpha -1)%
\right]  \notag \\
&&\times \left( 1-2\alpha +2\alpha ^{2}\right) ,
\end{eqnarray}%
where the functions $R_{i}(\alpha ,\beta )$ are:%
\begin{eqnarray}
&&R_{1}(\alpha ,\beta )=\beta ^{5}-4\beta ^{6}+6\beta ^{7}-4\beta ^{8}+\beta
^{9}+\beta ^{4}\alpha ^{5}+4\beta ^{3}\alpha ^{5}(\alpha -1)+6\beta
^{2}\alpha ^{5}(\alpha -1)^{2}+4\beta \alpha ^{5}(\alpha -1)^{3}  \notag \\
&&+\alpha ^{5}(\alpha -1)^{4};  \notag \\
&&R_{2}(\alpha ,\beta )=2\beta ^{9}+16\beta \alpha ^{6}(\alpha
-1)^{2}+2\alpha ^{7}(\alpha -1)^{2}+4\beta ^{8}(4\alpha -1)+\beta ^{5}\alpha
^{2}(12+64\alpha -111\alpha ^{2})  \notag \\
&&+\beta ^{6}\alpha (16-30\alpha -35\alpha ^{2})+\beta ^{3}\alpha
^{4}(-30+64\alpha -35\alpha ^{2})+6\beta ^{2}\alpha ^{5}(2-5\alpha +3\alpha
^{2})+2\beta ^{7}\left( 1-16\alpha \right.  \notag \\
&&\left. +9\alpha ^{2}\right) -3\beta ^{4}\alpha ^{3}(10-44\alpha +37\alpha
^{2});  \notag \\
&&R_{3}(\alpha ,\beta )=27\beta ^{13}+27\alpha ^{8}(\alpha -1)^{5}+75\beta
\alpha ^{7}(\alpha -1)^{4}(2\alpha -1)+15\beta ^{12}(-9+10\alpha )+\beta
^{11}\left( 270-675\alpha \right.  \notag \\
&&\left. +376\alpha ^{2}\right) +\beta ^{2}\alpha ^{6}(\alpha
-1)^{3}(42-333\alpha +376\alpha ^{2})+3\beta ^{10}(-90+400\alpha -487\alpha
^{2}+139\alpha ^{3})  \notag \\
&&+3\beta ^{3}\alpha ^{5}(\alpha -1)^{2}(25-23\alpha -130\alpha
^{2}+139\alpha ^{3})-\beta ^{4}\alpha ^{4}(\alpha -1)^{2}(-138+741\alpha
-1056\alpha ^{2}+326\alpha ^{3})  \notag \\
&&+\beta ^{9}(135-1050\alpha +2169\alpha ^{2}-1224\alpha ^{3}-326\alpha
^{4})+\beta ^{8}\left( -27+450\alpha -1501\alpha ^{2}+1128\alpha ^{3}\right.
\notag \\
&&\left. +1708\alpha ^{4}-1960\alpha ^{5}\right) +\beta ^{5}\alpha
^{3}(75-1017\alpha +4264\alpha ^{2}-7734\alpha ^{3}+6372\alpha
^{4}-1960\alpha ^{5})  \notag \\
&&-\beta ^{6}\alpha ^{2}(42+219\alpha -2676\alpha ^{2}+7734\alpha
^{3}-8756\alpha ^{4}+3438\alpha ^{5})  \notag \\
&&-\beta ^{7}\alpha (75-459\alpha +177\alpha ^{2}+3179\alpha ^{3}-6372\alpha
^{4}+3438\alpha ^{5});  \notag \\
&&R_{4}(\alpha ,\beta )=6\beta ^{13}+6\alpha ^{7}(\alpha -1)^{6}-12\beta
\alpha ^{6}(\alpha -1)^{5}(-6+5\alpha )-12\beta ^{12}(3+5\alpha )+\beta
^{11}(90+372\alpha -552\alpha ^{2})  \notag \\
&&-3\beta ^{2}\alpha ^{5}(\alpha -1)^{4}(75-257\alpha +184\alpha ^{2})-\beta
^{3}\alpha ^{4}(\alpha -1)^{3}(-273+1914\alpha -3696\alpha ^{2}+2087\alpha
^{3})  \notag \\
&&-\beta ^{10}(120+960\alpha -2979\alpha ^{2}+2087\alpha ^{3})+\beta
^{9}\left( 90+1320\alpha \right.  \notag \\
&&\left. -6621\alpha ^{2}+9957\alpha ^{3}-4948\alpha ^{4}\right) -\beta
^{4}\alpha ^{3}(\alpha -1)^{2}(273-2490\alpha +8327\alpha ^{2}-11038\alpha
^{3}+4948\alpha ^{4})  \notag \\
&&-\beta ^{5}\alpha ^{2}(\alpha -1)^{2}(225-2283\alpha +8789\alpha
^{2}-14735\alpha ^{3}+8390\alpha ^{4})-\beta ^{8}\left( 36+1020\alpha
-7734\alpha ^{2}+19263\alpha ^{3}\right.  \notag \\
&&\left. -20934\alpha ^{4}+8390\alpha ^{5}\right) +\beta ^{7}(6+420\alpha
-4986\alpha ^{2}+19190\alpha ^{3}-35351\alpha ^{4}+31515\alpha
^{5}-10793\alpha ^{6})  \notag \\
&&+\beta ^{6}\alpha (-72+1671\alpha -10257\alpha ^{2}+30182\alpha
^{3}-46649\alpha ^{4}+35918\alpha ^{5}-10793\alpha ^{6})  \notag \\
&&R_{5}(\alpha ,\beta )=99\beta ^{15}+99\alpha ^{7}(\alpha -1)^{6}+\beta
^{14}(-792+591\alpha )-3\beta \alpha ^{6}(\alpha -1)^{5}(116-263\alpha
+66\alpha ^{2})  \notag \\
&&+3\beta ^{13}(924-1495\alpha +615\alpha ^{2})+3\beta
^{12}(-1848+4949\alpha -4384\alpha ^{2}+1279\alpha ^{3})  \notag \\
&&+3\beta ^{2}\alpha ^{5}(\alpha -1)^{4}(154-926\alpha +1274\alpha
^{2}-460\alpha ^{3}+33\alpha ^{4})+\beta ^{11}\left( 6930-27993\alpha
+40629\alpha ^{2}\right.  \notag \\
&&\left. -25599\alpha ^{3}+6053\alpha ^{4}\right) +3\beta ^{3}\alpha
^{4}(\alpha -1)^{3}(-113+1295\alpha -3950\alpha ^{2}+4306\alpha
^{3}-1722\alpha ^{4}+199\alpha ^{5}) \\
&&+\beta ^{4}\alpha ^{3}(\alpha -1)^{3}(-339+3231\alpha -13948\alpha
^{2}+21305\alpha ^{3}-11307\alpha ^{4}+1809\alpha ^{5})  \notag \\
&&+\beta ^{10}(-5544+32865\alpha -70902\alpha ^{2}+73401\alpha
^{3}-37829\alpha ^{4}+7829\alpha ^{5})+\beta ^{5}\alpha ^{2}(\alpha
-1)^{2}\left( 462-3732\alpha \right.  \notag \\
&&\left. +15793\alpha ^{2}-38400\alpha ^{3}+43898\alpha ^{4}-21541\alpha
^{5}+3639\alpha ^{6}\right) +\beta ^{9}\left( 2772-24591\alpha +76245\alpha
^{2}\right.  \notag \\
&&\left. -117654\alpha ^{3}+100903\alpha ^{4}-46182\alpha ^{5}+8648\alpha
^{6}\right) +\beta ^{8}\left( -792+11445\alpha -51540\alpha
^{2}+114630\alpha ^{3}\right.  \notag \\
&&\left. -149274\alpha ^{4}+116108\alpha ^{5}-48984\alpha ^{6}+8414\alpha
^{7}\right) +\beta ^{6}\alpha \left( 348-4854\alpha +24609\alpha
^{2}-72793\alpha ^{3}\right.  \notag \\
&&\left. +137991\alpha ^{4}-162173\alpha ^{5}+110796\alpha ^{6}-39539\alpha
^{7}+5615\alpha ^{8}\right) +\beta ^{7}\left( 99-3027\alpha +21267\alpha
^{2}\right.  \notag \\
&&\left. -68907\alpha ^{3}+133130\alpha ^{4}-162246\alpha ^{5}+118955\alpha
^{6}-46584\alpha ^{7}+7313\alpha ^{8}\right) .
\end{eqnarray}

In expressions above, $\Theta (z)$ is Unit Step function. We have used also
the following short-hand notations%
\begin{eqnarray}
N_{1} &=&\beta ^{2}+\beta (\alpha -1)+\alpha (\alpha -1),\ \ \ \ \
N_{2}=(\alpha +\beta )N_{1},\ L=\alpha +\beta -1,\   \notag \\
,\ \ \ L_{1} &\equiv &L_{1}(s,\alpha ,\beta )=\frac{(1-\beta )}{N_{1}^{2}}%
\left[ m_{c}^{2}N_{2}-s\alpha \beta L\right] ,\ L_{2}\equiv L_{2}(s,\alpha
)=s\alpha (1-\alpha )-m_{c}^{2},\
\end{eqnarray}

\end{widetext}


\begin{thebibliography}{99}

\bibitem{Aaij:2021vvq} R.~Aaij \textit{et al.} [LHCb Collaboration],
arXiv:2109.01038 [hep-ex].


\bibitem{LHCb:2021auc} R.~Aaij \textit{et al.} [LHCb Collaboration],
arXiv:2109.01056 [hep-ex].


\bibitem{Agaev:2021vur} S.~S.~Agaev, K.~Azizi, and H.~Sundu,
Nucl.\ Phys.\ B \textbf{975}, 115650 (2022). 


\bibitem{Feijoo:2021ppq} A.~Feijoo, W.~H.~Liang, and E.~Oset,
arXiv:2108.02730 [hep-ph].


\bibitem{Yan:2021wdl} M.~J.~Yan, and M.~P.~Valderrama, Phys. Rev. D \textbf{%
105}, 014007 (2022). 


\bibitem{Fleming:2021wmk} S.~Fleming, R.~Hodges, and T.~Mehen, Phys. Rev. D
\textbf{104}, 116010 (2021). 


\bibitem{Azizi:2021aib} K.~Azizi, U.~\"{O}zdem, Phys. Rev. D \textbf{104},
114002 (2021).


\bibitem{Meng:2021jnw} L.~Meng, G.~J.~Wang, B~.Wang, and S.~L.~Zhu, Phys.\
Rev.\ D \textbf{104}, 051502 (2021). 


\bibitem{Ling:2021bir} X.~Z.~Ling, M.~Z.~Liu, L.~S~.Geng, E.~Wang, and
J.~J.~Xie, arXiv:2108.00947 [hep-ph].


\bibitem{Chen:2021vhg} R.~Chen, Q.~Huang, X.~Liu, and S.~L.~Zhu, Phys.\
Rev.\ D \textbf{104}, 114042 (2021). 


\bibitem{Xin:2021wcr} Q.~Xin, and Z.~G.~Wang,
arXiv:2108.12597 [hep-ph].


\bibitem{Agaev:2018khe} S.~S.~Agaev, K.~Azizi, B.~Barsbay, and H.~Sundu,
Phys.\ Rev.\ D \textbf{99}, 033002 (2019).


\bibitem{Agaev:2019lwh} S.~S.~Agaev, K.~Azizi, B.~Barsbay, and H.~Sundu,
Phys.\ Rev.\ D \textbf{101}, 094026 (2020). 


\bibitem{Agaev:2020zag} S.~S.~Agaev, K.~Azizi, B.~Barsbay, and H.~Sundu,
Chin.\ Phys.\ C \textbf{45}, 013105 (2021). 


\bibitem{Agaev:2019kkz} S.~S.~Agaev, K.~Azizi and H.~Sundu,
Nucl.\ Phys.\ B \textbf{951}, 114890 (2020).


\bibitem{Karliner:2017qjm} M.~Karliner and J.~L.~Rosner,
Phys.\ Rev.\ Lett.\ \textbf{119}, 202001 (2017).


\bibitem{Eichten:2017ffp} E.~J.~Eichten and C.~Quigg,
Phys.\ Rev.\ Lett.\ \textbf{119}, 202002 (2017).


\bibitem{Agaev:2020zad} S.~S.~Agaev, K.~Azizi and H.~Sundu,
Turk.\ J.\ Phys.\ \textbf{44}, 95 (2020). 


\bibitem{Navarra:2007yw} F.~S.~Navarra, M.~Nielsen and S.~H.~Lee,
Phys.\ Lett.\ B \textbf{649}, 166 (2007).


\bibitem{Du:2012wp} M.~L.~Du, W.~Chen, X.~L.~Chen and S.~L.~Zhu,
Phys.\ Rev.\ D \textbf{87}, 014003 (2013).


\bibitem{Aaij:2017ueg} R.~Aaij \textit{et al.} [LHCb Collaboration],
Phys.\ Rev.\ Lett.\ \textbf{119}, 112001 (2017).


\bibitem{Wang:2017dtg} Z.~G.~Wang, and Z.~H.~Yan,
Eur.\ Phys.\ J.\ C \textbf{78}, 19 (2018). 


\bibitem{Braaten:2020nwp} E.~Braaten, L.~P.~He, and A.~Mohapatra,
Phys.\ Rev.\ D \textbf{103}, 016001 (2021). 


\bibitem{Cheng:2020wxa} J.~B.~Cheng, S.~Y.~Li, Y.~R.~Liu, Z.~G.~Si, and
T.~Yao, 
Chin.\ Phys.\ C \textbf{45}, 043102 (2021). 


\bibitem{Meng:2020knc} Q.~Meng, E.~Hiyama, A.~Hosaka, M.~Oka, P.~Gubler,
K.~U.~Can, T.~T.~Takahashi, and H.~S.~Zong,
Phys.\ Lett.\ B \textbf{814}, 136095 (2021). 


\bibitem{Junnarkar:2018twb} P.~Junnarkar, N.~Mathur, and M.~Padmanath,
Phys.\ Rev.\ D \textbf{99}, 034507 (2019). 


\bibitem{Agaev:2019qqn} S.~S.~Agaev, K.~Azizi, and H.~Sundu,
Phys.\ Rev.\ D \textbf{99}, 114016 (2019). 


\bibitem{Agaev:2018vag} S.~S.~Agaev, K.~Azizi, B.~Barsbay and H.~Sundu,
Nucl.\ Phys.\ B \textbf{939}, 130 (2019).


\bibitem{Novikov:1977dq} V.~A.~Novikov, L.~B.~Okun, M.~A.~Shifman,
A.~I.~Vainshtein, M.~B.~Voloshin, and V.~I.~Zakharov, Phys.\ Rept.\ \textbf{%
41}, 1 (1978). 


\bibitem{Dias:2011mi} J.~M.~Dias, S.~Narison, F.~S.~Navarra, M.~Nielsen, and
J.~M.~Richard, 
Phys.\ Lett.\ B \textbf{703}, 274 (2011). 


\bibitem{Li:2012ss} N.~Li, Z.~F.~Sun, X.~Liu and S.~L.~Zhu,
Phys.\ Rev.\ D \textbf{88}, 114008 (2013). 


\bibitem{Shifman:1978bx} M.~A.~Shifman, A.~I.~Vainshtein and V.~I.~Zakharov,
Nucl.\ Phys.\ B \textbf{147}, 385 (1979).


\bibitem{Shifman:1978by} M.~A.~Shifman, A.~I.~Vainshtein and V.~I.~Zakharov,
Nucl.\ Phys.\ B \textbf{147}, 448 (1979).


\bibitem{Kondo:2004cr} Y.~Kondo, O.~Morimatsu and T.~Nishikawa,
Phys.\ Lett.\ B \textbf{611}, 93 (2005).


\bibitem{Lee:2004xk} S.~H.~Lee, H.~Kim and Y.~Kwon,
Phys.\ Lett.\ B \textbf{609}, 252 (2005).


\bibitem{Wang:2015nwa} Z.~G.~Wang,
Int.\ J.\ Mod.\ Phys.\ A \textbf{30}, 1550168 (2015).


\bibitem{Sundu:2018nxt} H.~Sundu, S.~S.~Agaev and K.~Azizi, Eur.\ Phys.\ J.
C 79, 215 (2019). 


\bibitem{Maiani:2014} L.~Maiani, F.~Piccinini, A.~D.~Polosa and V.~Riquer,
Phys.\ Rev.\ D \textbf{89}, 114010 (2014).


\bibitem{Wang:2014vha} Z.~G.~Wang,
Commun.\ Theor.\ Phys.\ \textbf{63}, 325 (2015).


\bibitem{Agaev:2017tzv} S.~S.~Agaev, K.~Azizi and H.~Sundu,
Phys.\ Rev.\ D \textbf{96}, 034026 (2017).


\bibitem{Kim:2022mpa} Y.~Kim, M.~Oka and K.~Suzuki,
Phys.\ Rev.\ D \textbf{105}, 074021 (2022). 


\bibitem{PDG:2020} P.~A.~Zyla \textit{et al.} [Particle Data Group], Prog.\
Theor.\ Exp.\ Phys.\ \textbf{2020}, 083C01 (2020).


\bibitem{Agaev:2020nrc} S.~S.~Agaev, K.~Azizi and H.~Sundu,
J.\ Phys.\ G. \textbf{48}, 085012 (2021).


\bibitem{He:2020jna} X.~G.~He, W.~Wang and R.~Zhu,
Eur.\ Phys.\ J.\ C \textbf{80}, 1026 (2020). 


\bibitem{Albuquerque:2022weq} R.~Albuquerque, S.~Narison, and
D.~Rabetiarivony 
Nucl.\ Phys.\ A \textbf{1023}, 122451 (2022). 
\end{thebibliography}
\end{document}